\documentclass{aa}
\usepackage{graphicx}

\begin{document}

\title{Analysis of the distribution of HII regions in external galaxies}
\subtitle{IV The new galaxy sample. Position and inclination angles}
\author{C. Garc\'{\i}a-G\'omez\inst{1} \and E. Athanassoula \inst{2}
\and C. Barber\`a\inst{1}}
\institute{D.E.I.M., Campus Sescelades, Avd. dels Pa\"{\i}sos
Catalans 26, 43007 \\ Tarragona, Spain
\and Observatoire de Marseille, 2 Place Le Verier, 13248 Marseille
cedex 04, France}
\offprints{C. Garc\'{\i}a-G\'omez, \email{cgarcia@etse.urv.es}}
\date{Received / Accepted}
\abstract{We have compiled a new sample of galaxies with published
catalogs of HII region coordinates. This sample, together with the
former catalog of Garc\'{\i}a-G\'omez and Athanassoula (\cite{gga1}),
will form the basis for subsequent studies of the spiral structure in
disc galaxies. In this paper we address the problem of the
deprojection of the galaxy images. For this purpose we use two deprojection
methods based on the HII region distribution and compare the results
with the values found in the literature using other deprojection
methods. Taking into account the results of all the methods, we
propose optimum values for the position and inclination angles of all
 the galaxies in our sample.
\keywords{galaxies--structure--spiral galaxies--interstellar
medium:HII regions}
}
\titlerunning{Analysis of the distribution of HII regions}
\maketitle

\section{Introduction}

Catalogs of HII regions of spiral galaxies can be used to obtain
a great deal of information on their underlying galaxies. For
instance, we can study the radial profile of the distribution of HII
regions and compare it with the radial distribution of the light in
different wavelengths. We
can study also the number distribution of HII regions according to
their flux and use this information to study the star formation rate
in galaxies. Yet more information can be obtained by Fourier
analyzing the spatial distribution of the HII regions. The
distribution can then be decomposed into components of a given angular
periodicity, whose pitch angles and relative amplitude can be
calculated. Using such results for a large sample of galaxies 
should provide useful information on the properties of spiral structure in
external galaxies. With this aim, we started in 1991 a study of all
catalogs published until that date (Garc\'{\i}a-G\'omez \&
Athanassoula \cite{gga1}, hereafter GGA; Garc\'{\i}a-G\'omez \&
Athanassoula \cite{gga2};
Athanassoula, Garc\'{\i}a-G\'omez \& Bosma
\cite{athanassoula}). Since then,
however, a large number of new catalogs has been published,
 nearly doubling our initial sample. This warrants a reanalysis of
our previous results, particularly since the new
published catalogs extend the initial sample specially towards barred
and ringed galaxies, which were sparse in the first sample. The new
sample contains also a large number of active galaxies.

In this paper we will concentrate on the fundamental step of the
deprojection of the HII region positions from the plane of the sky onto the
plane of the galaxy, determining the position angle (hereafter PA) and
the inclination angle (hereafter IA) of each galaxy using several
methods of deprojection. In future papers we will study the spiral
structure of the galaxies which have a sufficient number of HII
regions and use the whole 
sample to study the global properties of the spiral structure in
external galaxies. This paper is structured as follows: In Section~2 we
present the new sample and in Section~3 we discuss the different
methods of deprojection used to obtain the values of PA and IA. In
Section~4 we give the PA and IA values adopted for the galaxies in our
sample and finally in Section~5 we compare the values of the PA and IA
obtained by using the different methods of deprojection.

\section{The sample}

Several new catalogs of the distribution of HII regions in external
galaxies have been published in the last decade. These come
mainly from the surveys of H$\alpha$ emission in Seyfert galaxies
from Tsvetanov \& Petrosian (\cite{tsvetanov}), Evans et
al. (\cite{evans}) and
Gonz\'alez Delgado et al. (\cite{gonzalez}), from the H$\alpha$
survey of southern galaxies from Feinstein (\cite{feinstein})
and the H$\alpha$ survey of ringed galaxies by Crocker et al.
(\cite{crocker}). There have also been a considerable
number of studies of the HII region distribution in individual
galaxies: \object{NGC~157}, \object{NGC~3631}, \object{NGC~6764} and
\object{NGC~6951} (Rozas et al. \cite{rozas1}); \object{NGC~598} 
(Hodge et al. \cite{hodge2}):
\object{NGC~3198} (Corradi et al. \cite{corradi});
\object{NGC~3359}
(Rozas et al. \cite{rozas3}); \object{NGC~4321} (Knapen \cite{knapen2});
\object{NGC~4258} (Court\`es et al. \cite{courtes});
\object{NGC 4736} (Rodrigues et al. \cite{rodrigues});
\object{NGC~5194} (Petit et al. \cite{petit2});
\object{NGC~5457} (Hodge et al. \cite{hodge}, Scowen et al. \cite{scowen});
\object{NGC~6184} (Knapen et al. \cite{knapen3}); \object{NGC 7331}
(Petit \cite{petit1}) and \object{NGC~7479} (Rozas et al. \cite{rozas4}).

\begin{figure}{t}
\includegraphics[angle=-90,scale=0.37]{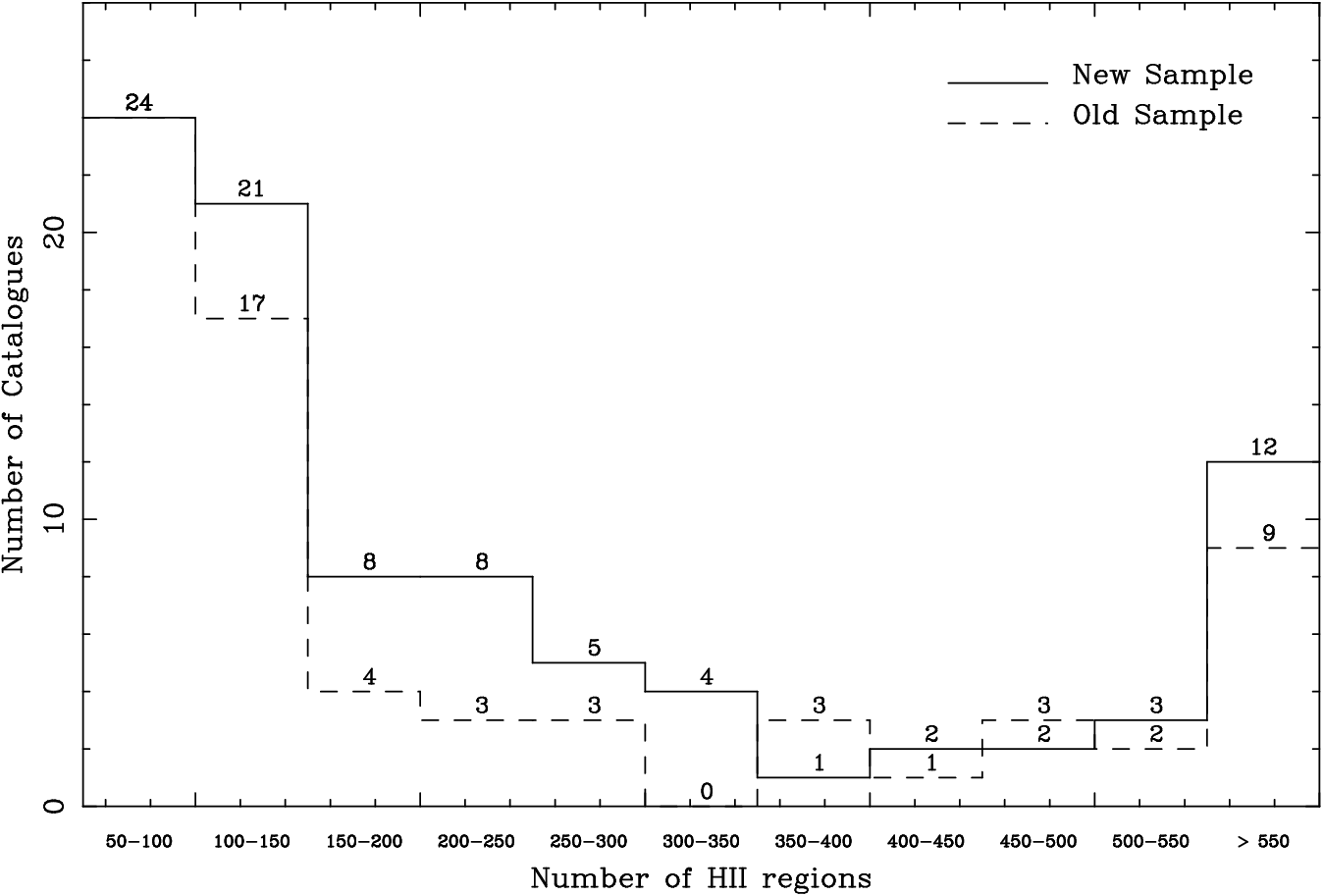}
\caption{Distribution of galaxies as a function of number of HII regions.}
\label{figHIIreg}
\end{figure}
\vfill

\begin{figure}
\includegraphics[angle=-90,scale=0.42]{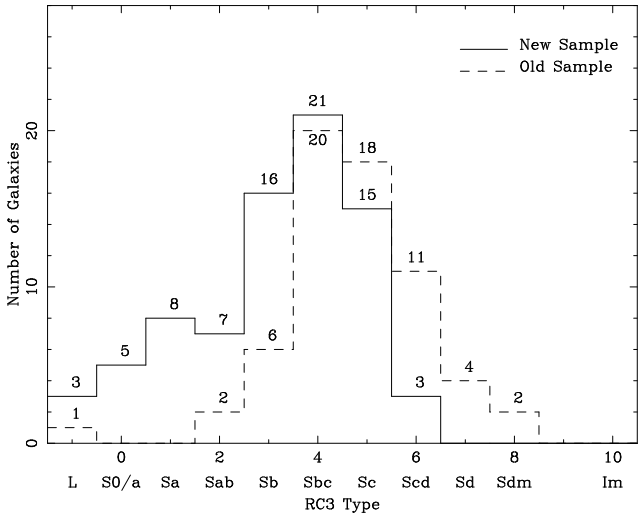}
\caption{Distribution of galaxies as a function of galaxy type.}
\label{figtype}
\end{figure}
\vfill
From these studies we have retained all galaxies with an
inclination less than $80^o$ and with a sufficient number of HII regions
listed. Indeed, more than $50$ HII regions are necessary to
allow a study of the spiral structure.
Nevertheless, we included also five galaxies with fewer
HII regions because we judged that they described fairly
well the galaxy disc. In Fig~\ref{figHIIreg} we show the number of
catalogs as a function of the number of HII regions they have. The
solid line corresponds to catalogs of the new sample, and the dashed line
corresponds to the sample used in GGA (\cite{gga1}). From this figure we see that
the quality of the two samples is roughly the same.
\vfill

\begin{center}
\begin{table}[ht]
\caption{General Properties of the galaxies in the new sample.}
\label{hiicat}
\begin{tabular}{|lrrrrrr|}
\hline
Name &  T &  S  &  AC  &  SS  &  N  &  Ref.\\\hline
ESO 111-10& PSXT4..& 3.7& - & N &  64& 1\\
ESO 152-26& PSBR1..& 1.0& - & Y & 236& 1\\ 
ESO 377-24& .SAT5*P& 5.0& - & N &  59& 2\\ 
IC 1438   & PSXT1*.& 0.7& - & Y &  44& 1\\ 
IC 2510   & .SBT2*.& 2.3& - & N &  70& 2\\
IC 2560   & PSBR3*.& 3.3& - & R & 137& 2\\ 
IC 3639   & .SBT4*.& 4.0& - & N & 112& 2\\ 
IC 4754   & PSBR3*.& 2.6& - & N & 114& 1\\ 
IC 5240   & .SBR1..& 1.0& - & N & 119& 1\\ 
NGC 0053  & PSBR3..& 0.6& - & N &  66& 1\\ 
NGC 0157  & .SXT4..& 4.0& 12& Y & 707& 8\\ 
NGC 0210  & .SXS3..& 3.0& 6 & Y & 518& 1\\ 
NGC 0598  & .SAS6..& 6.0& 5 & N &1272&18\\
NGC 1068  & RSAT3..& 3.0& 3 & N & 110& 3\\ 
NGC 1097  & .SBS3..& 3.0& 12& Y & 401& 6\\ 
NGC 1386  & .LBS+..&-0.6& - & N &  44& 2\\ 
NGC 1433  & PSBR2..& 2.0& 6 & R & 779& 1\\
NGC 1566  & .SXS4..& 4.0& 12& Y & 679& 2\\ 
NGC 1667  & .SXR5..& 5.0& - & N &  46& 3\\ 
NGC 1672  & .SBS3..& 3.0& 5 & N & 260& 6\\
NGC 1808  & RSXS1..& 1.0& - & N & 206& 2\\ 
NGC 1832  & .SBR4..& 4.0& 5 & R & 206& 1\\ 
NGC 2985  & PSAT2..& 2.0& 3 & N & 110& 3\\ 
NGC 2997  & .SXT5..& 5.0& 9 & R & 373& 5\\ 
NGC 3081  & RSXR0..& 0.0& 6 & N &  75& 6\\ 
          &        &    &   & N &  58& 2\\ 
NGC 3198  & .SBT5..& 5.0& - & N & 104& 9\\ 
NGC 3359  & .SBT5..& 5.0& 5 & Y & 547& 4\\ 
NGC 3367  & .SBT5..& 5.0& 9 & R &  79& 3\\ 
NGC 3393  & PSBT1*.& 1.0& - & R &  80& 2\\ 
NGC 3631  & .SAS5..& 5.0& 9 & Y &1322& 8\\ 
NGC 3660  & .SBR4..& 4.0& 2 & N &  59& 3\\ 
NGC 3783  & PSBR2..& 1.5& 9 & N &  58& 2\\ 
NGC 3982  & .SXR3*.& 4.6& 2 & N & 117& 3\\
NGC 4051  & .SXT4..& 4.0& 5 & N & 123& 6\\
NGC 4123  & .SBR5..& 5.0& 9 & N &  58& 5\\
NGC 4258  & .SXS4..& 4.0& - & Y & 136& 16\\
NGC 4321  & .SXS4..& 4.0& 12& Y &1948& 12\\
NGC 4507  & PSXT3..& 3.0& 5 & N &  92& 2\\
NGC 4593  & RSBT3..& 3.0& 5 & Y & 112& 2 \\
          &        &    &   & R &  45& 5\\
NGC 4602  & .SXT4..& 4.0& - & N & 218& 2 \\
          &        &    & - & N &  46& 5\\
NGC 4639  & .SXT4..& 4.0& 2 & R & 190& 6 \\
NGC 4699  & .SXT3..& 3.0& 3 & N & 104& 3\\
NGC 4736  & RSAR2..& 2.0& 3 & N & 168& 3 \\
          &        &    &   & N &  90& 17 \\
NGC 4939  & .SAS4..& 4.0& 12& Y & 250& 2\\
          &        &    &   & Y & 206& 6 \\
NGC 4995  & .SXT3..& 3.0& 6 & R & 142& 3\\
NGC 5033  & .SAS5..& 5.0& 9 & R & 423& 6\\
NGC 5194  & .SAS4P.& 4.0& 12& Y & 477& 7\\
NGC 5364  & .SAT4P.& 4.0& 9 & R & 174& 5\\
NGC 5371  & .SXT4..& 4.0& 9 & Y & 100& 3\\
\hline
\end{tabular}
\end{table}
\vfill

\begin{table}[t]
\addtocounter{table}{-1}
\caption{Continued}
\begin{tabular}{|lrrrrrr|}
\hline
Name &  T &  S  &  AC  &  SS  &  N  &  Ref.\\\hline
NGC 5427  & .SAS5P.& 5.0& 9 & Y & 300& 2\\
          &        &    &   & Y & 164& 1\\
          &        &    &   & R &  78& 6\\
NGC 5457  & .SXT6..& 6.0& 9 & Y &1264& 13\\
          &        &    &   & Y & 248& 14\\
NGC 5643  & .SXT5..& 5.0& - & N & 214& 2\\
NGC 5861  & .SXT5..& 5.0& 12& N &  55& 5\\
NGC 6070  & .SAS6..& 6.0& 9 & N &  61& 5\\
NGC 6118  & .SAS6..& 6.0& - & N & 117& 5 \\
NGC 6221  & .SBS5..& 5.0& - & Y & 173& 2\\
NGC 6300  & .SBT3..& 3.0& 6 & N & 977& 1\\
          &        &    &   & N & 317& 6\\
NGC 6384  & .SXR4..& 4.0& 9 & R & 283& 5 \\
NGC 6753  & RSAR3..& 3.0& 8 & Y & 541& 1 \\
NGC 6764  & .SBS4..& 3.5& 5 & R & 348& 8\\
NGC 6782  & RSXR1..& 0.8& - & Y & 296& 1 \\
NGC 6814  & .SXT4..& 4.0& 9 & Y & 734& 15\\
          &        &    &   & Y & 131& 6\\
NGC 6902  & .SAR3..& 3.1& - & R & 467& 1\\
NGC 6935  & PSAR1..& 1.1& - & N & 166& 1\\
NGC 6937  & PSBR5*.& 4.9& - & Y & 213& 1\\ 
NGC 6951  & .SXT4..& 4.0& 12& Y & 664& 8\\
NGC 7020  & RLAR+..&-1.0& - & N &  68& 1\\
NGC 7098  & RSXT1..& 1.0& - & R & 188& 1\\
NGC 7219  & RSAR0P.& 0.0& - & Y & 139& 1\\
NGC 7267  & PSBT1P.& 1.0& - & N & 122& 1\\
NGC 7314  & .SXT4..& 4.0& 2 & N & 151& 6\\
          &        &    &   & N & 117& 2\\
NGC 7329  & .SBR3..& 3.0& - & R & 349& 1\\
NGC 7331  & .SAS3..& 3.0& 3 & N & 252& 10\\
NGC 7479  & .SBS5..& 5.0& 9 & R & 126& 11\\
NGC 7531  & .SXR4..& 2.8& 3 & R & 549& 1\\
NGC 7552  & PSBS2..& 2.0& - & N &  78& 5\\
NGC 7590  & .SAT4*.& 4.0& - & N & 129& 2\\\hline
\multicolumn{7}{|l|}{(1) Crocker, D.A., et al. \cite{crocker}} \\
\multicolumn{7}{|l|}{(2) Tsvetanov, Z.I., Petrosian, A.R. \cite{tsvetanov}}\\
\multicolumn{7}{|l|}{(3) Gonz\'alez Delgado, R.M., et al. \cite{gonzalez}}\\
\multicolumn{7}{|l|}{(4) Rozas, M., et al. \cite{rozas2}}\\
\multicolumn{7}{|l|}{(5) Feinstein, C. \cite{feinstein}}\\
\multicolumn{7}{|l|}{(6) Evans, I.N., et al. \cite{evans}}\\
\multicolumn{7}{|l|}{(7) Petit, H., et al. \cite{petit2}}\\
\multicolumn{7}{|l|}{(8) Rozas, M., et al. \cite{rozas1}}\\
\multicolumn{7}{|l|}{(9) Corradi, R.L.M. et al. \cite{corradi}}\\
\multicolumn{7}{|l|}{(10) Petit, H. \cite{petit1}}\\
\multicolumn{7}{|l|}{(11) Rozas, M., et al. \cite{rozas4}}\\
\multicolumn{7}{|l|}{(12) Knapen, J.H. \cite{knapen2}}\\
\multicolumn{7}{|l|}{(13) Hodge, P.W., et al. \cite{hodge}}\\
\multicolumn{7}{|l|}{(14) Scowen, P., et al. \cite{scowen}}\\
\multicolumn{7}{|l|}{(15) Knapen, J.H., et al. \cite{knapen3}}\\
\multicolumn{7}{|l|}{(16) Court\'es, G., et al. \cite{courtes}}\\
\multicolumn{7}{|l|}{(17) Rodrigues, I., et al. \cite{rodrigues}}\\
\multicolumn{7}{|l|}{(18) Hodge, P.W., et al \cite{hodge2}}\\\hline
\end{tabular}
\end{table}
\end{center}

\begin{figure}[ht]
\includegraphics[scale=0.5]{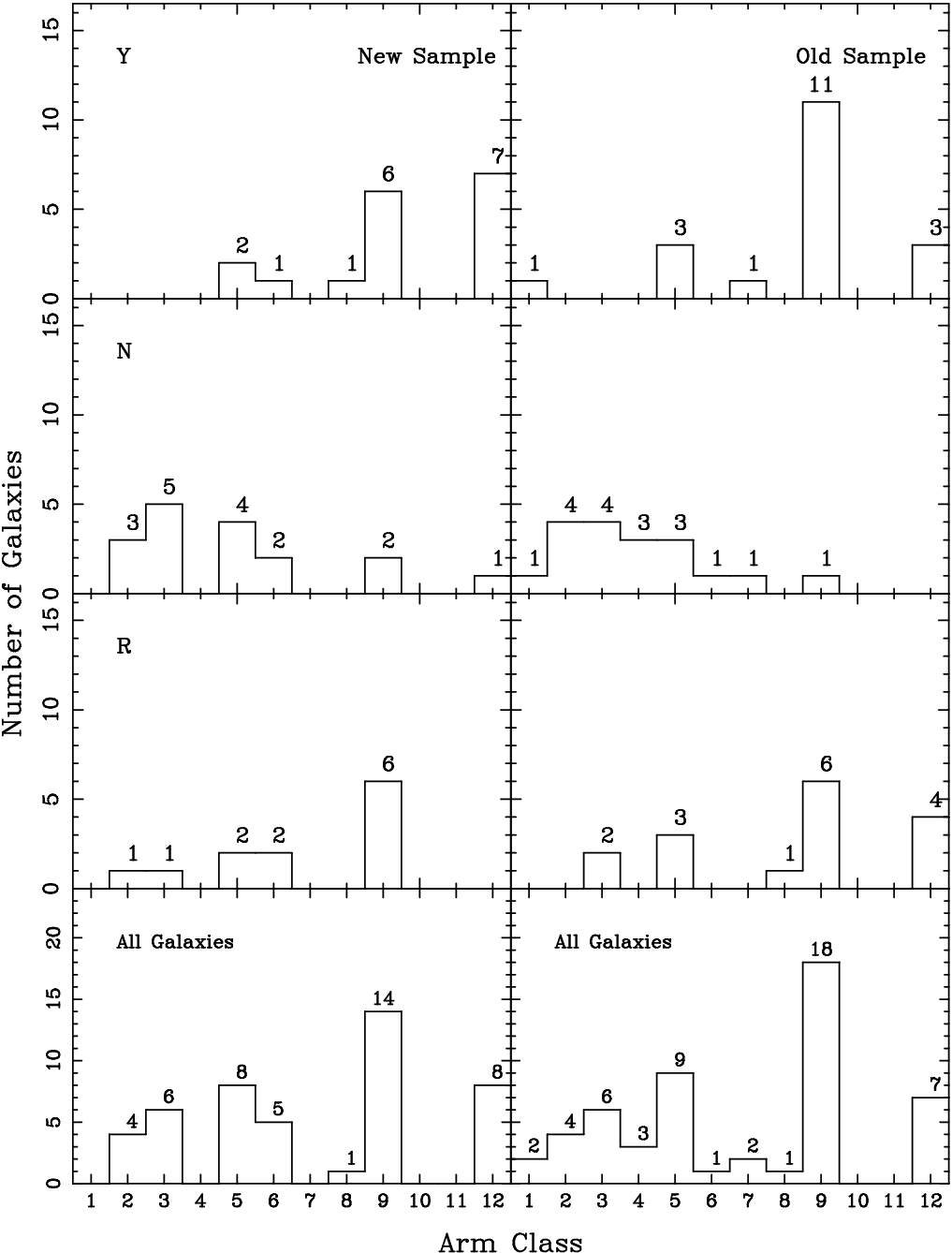}
\caption{Comparison of the arm classification of Elmegreen \&
 Elmegreen (\cite{elmegreen}) with our own classification of the arm
 structure present in the HII region distribution. The histograms give
 the number of galaxies as a function of the Elmegreen \& Elmegreen
 (\cite{elmegreen}) arm class. Left panels are for the new sample and
 right ones for the old one. The upper panels contain galaxies with
 prominent spiral structure in their HII region distribution. The
 second row, galaxies with no such structure. The third row contains
 galaxies which are intermediate or unclassifiable, and the fourth
 row, all galaxies together.}
\label{figmor}
\end{figure}

The properties of the galaxies selected for our
study and of their HII region catalogs are listed in
Table~\ref{hiicat}. Column~1 gives the galaxy name. Columns 2 and 3 the Hubble
type (T) and the Hubble stage (S) from the RC3 catalog (de Vaucouleurs
et al. \cite{devaucouleurs}).
 Column 4 gives the arm class (AC) as defined
in Elmegreen \& Elmegreen (\cite{elmegreen}). Column 5 classifies
the spiral structure (SS) seen in the

HII region distribution, which we
obtain by eye estimates as in GGA (\cite{gga1}).
If we can see a clear grand design spiral we give a
classification Y in Column 5 and N on the contrary. Intermediate cases
are classified as R. Finally, in Column 6 we give
the  number of HII regions in the catalog (N) and in Column 7 a key for
the  reference of this particular catalog. 
The number distribution of galaxies as a function of galaxy type is
shown in Fig~\ref{figtype}. The solid line corresponds to the
number distribution of types for the galaxies of the new sample, while
the dashed line corresponds to the sample used in GGA (\cite{gga1}). As in
the previous paper the peak is for galaxies of type Sbc. Nevertheless,
this sample contains many more earlier type galaxies than the previous
one. Indeed, in the previous sample most galaxies are of type Sbc and
later, while in the new one most are Sbc and earlier.

\begin{figure}
\includegraphics[scale=0.5]{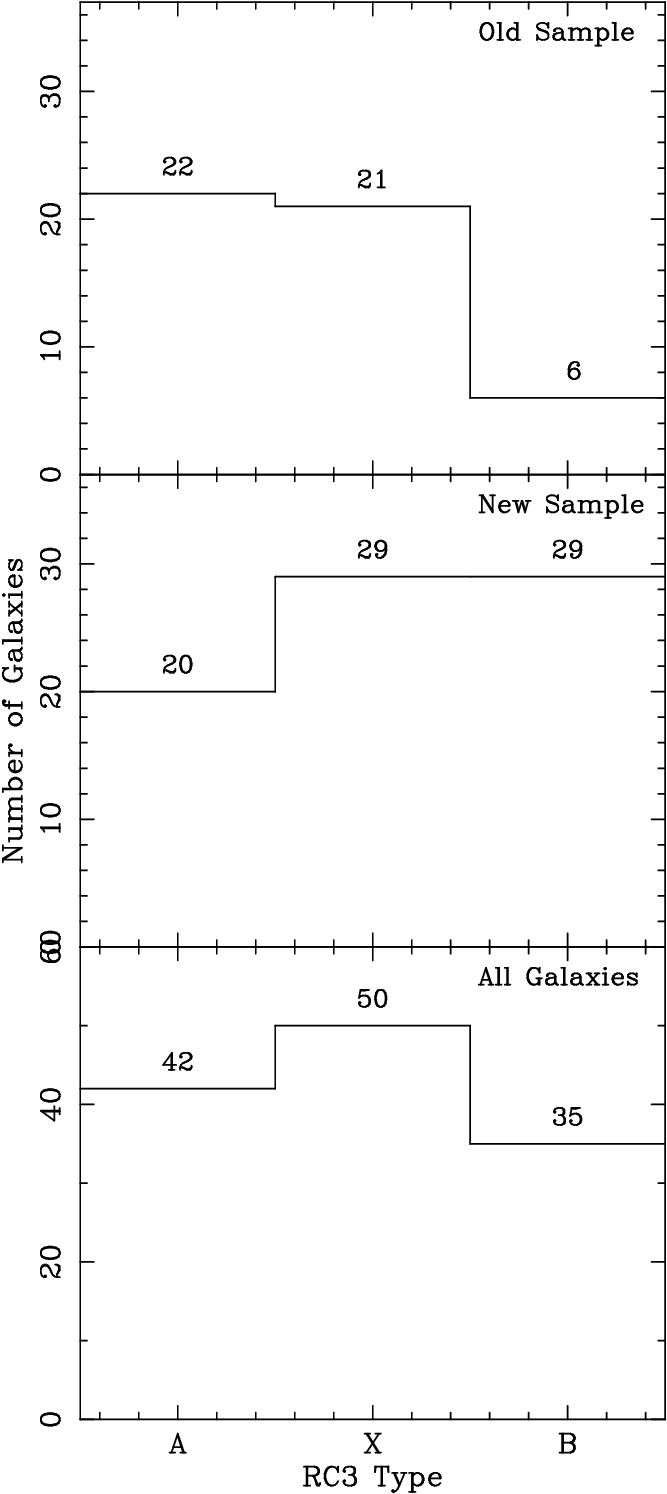}
\caption{Distribution of galaxies according to the bar type
classification of the RC3 (\cite{devaucouleurs}) catalog.}
\label{figbar}
\end{figure}

In Fig~\ref{figmor} we
compare the arm class classification of Elmegreen \& Elmegreen
(\cite{elmegreen}) with our classification of the arm structure
present in the HII region distribution. The left panels refer to the
galaxies in the new sample and the right panels to the galaxies in
the GGA (\cite{gga1}) sample. The upper panels correspond to
galaxies classified as Y, the
second row of panels correspond to the galaxies classified as N
and the third row contains the galaxies classified as
R. The last row contains the distribution of all the galaxies
according to arm class. The figure shows that galaxies
classified with higher number in the arm class have a tendency to have
well developed arms in the distribution of HII regions. There are,
however, exceptions as e.g. \object{NGC~5861} which has been given an
arm class 12 by Elmegreen \& Elmegreen (\cite{elmegreen}), but has no
apparent spiral structure in the HII region distribution.

We should also note that our new sample is formed
mainly by barred and ringed galaxies. This is shown in
Fig~\ref{figbar} where we can see the number 
distribution of galaxies as a function of bar type as given in the
RC3 catalog. In the upper panel we show the distribution of bar
types of the galaxies of the GGA (\cite{gga1}) sample.
In the middle panel we show the distribution for the
new sample and in the lower panel the distribution of bar types of 
the two samples combined. While the first sample was biased towards
non-barred galaxies (upper panel), the new catalog is biased towards
barred and/or ringed galaxies (middle panel). This gives a quite
uniform total distribution (lower panel).

\begin{figure}
\includegraphics[scale=0.45]{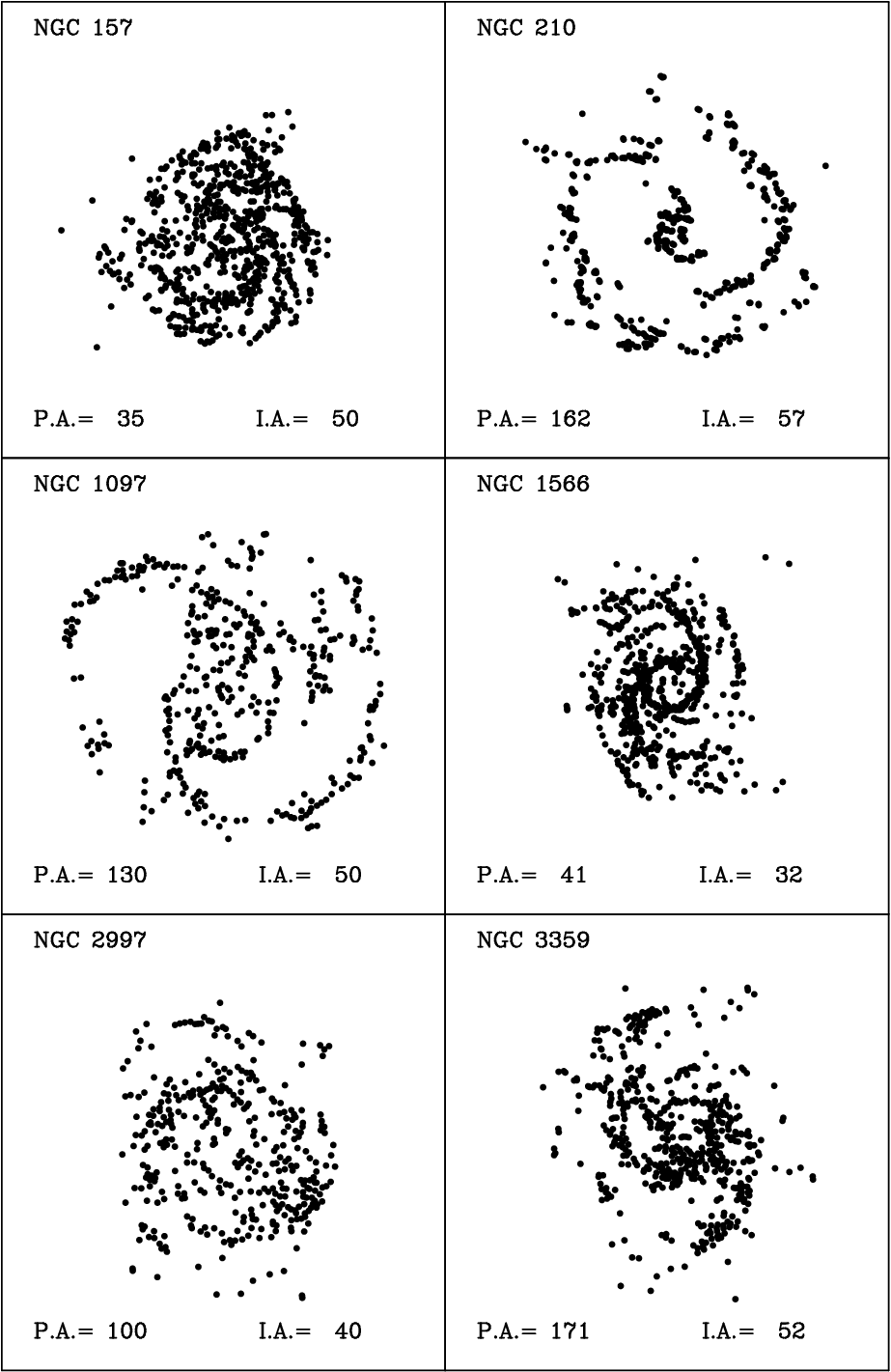}
\caption{Deprojected HII region distribution of the galaxies in our
sample with rich catalogs and clear spiral structure.}
\label{album1}
\end{figure}

\begin{figure}
\addtocounter{figure}{-1}
\includegraphics[scale=0.45]{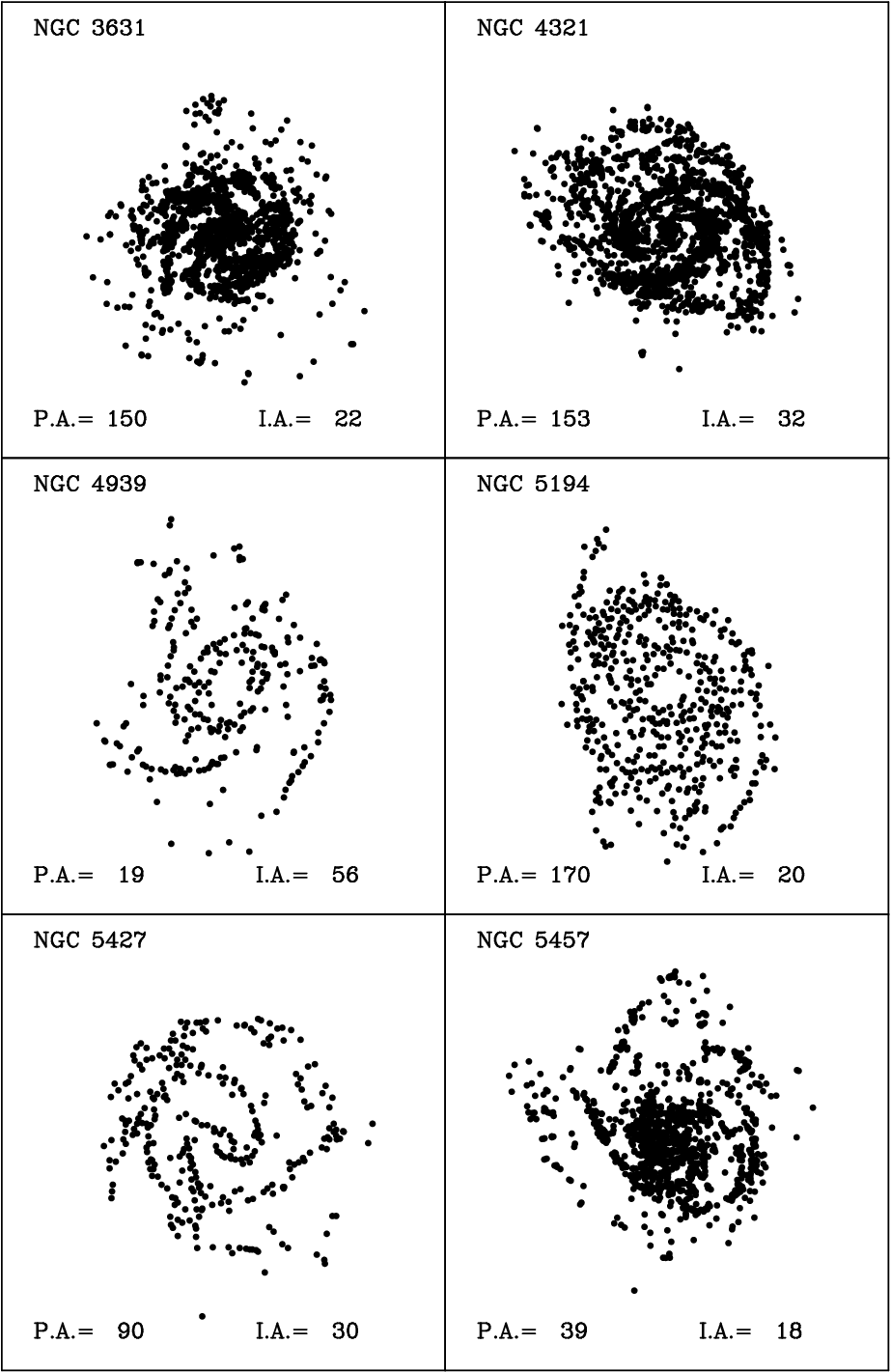}
\caption{Continued.}
\end{figure}

Our final aim is to study the spiral structure outlined by the HII
region distribution. Obviously, this cannot be done with a catalog
with a number of HII regions lower than 50, and only the richer
catalogs will be considered for this purpose in a forthcoming
paper. Nevertheless, useful information on the galaxies
can be obtained even from the less rich catalogs, as e.g. the
 orientation of the galaxies as seen in the sky and
the radial scale of the distribution, which can be compared with the
radial scale length of galaxy discs (Athanassoula et al. \cite{athanassoula}).

\section{Deprojection methods.}

The first step in order to study the distribution of HII regions is to
deproject the images of the galaxies. For this we need two angles, namely 
the position angle
(PA) -- which is the angle between the line of nodes of the
projected image and the north, measured towards the east -- and 
the inclination angle (IA) -- which is the angle between the
line perpendicular to the plane of the galaxy and the line of
sight --. An IA of zero degrees corresponds to a galaxy seen face-on.

Two basic groups of methods have been used so far to determine these
angles. The first one is based on photometry and images, and the
second one on kinematics. The most standard way to use
images is to fit ellipses to the outermost isophotes and measure their
axial ratio. This method, which is often used in the literature, is
well suited for discs which are not warped, and requires photometric
images with high
signal-to-noise ratio in the outer parts, where the influence of non
axisymmetric components like arms and bars is minimum.
Several variants have also been proposed: Danver (\cite{danver}) used
a special display table to rotate the galaxy images until they were
circular. Grosb\o l (\cite{grosbol1}) applied the
one dimensional Fourier transform to the intensity distribution in the
outer parts of disc galaxies and adopted the deprojection angles that
minimized the bisymmetric Fourier component.

Another classical method uses a two dimensional
velocity field of the galaxy. Assuming that the emission comes from a
thin planar disk, which is in circular motion around the galaxy center,
we can select the deprojection angles that minimize the departures
from such a flow. This method is particularly well adapted for HI
kinematics, which covers the
whole galaxy disk. During the last decade, however, this method has
been applied also to data coming from CO and H$\alpha$ kinematics, which
are generally more restricted to the central parts. This method is
specially well suited for measuring the PA. In the case of warps, a
tilted ring model can be used, but the
values of the deprojection angles will not be uniquely defined, as they
will change with galactocentric radius.

\begin{figure}
\addtocounter{figure}{-1}
\includegraphics[scale=0.45]{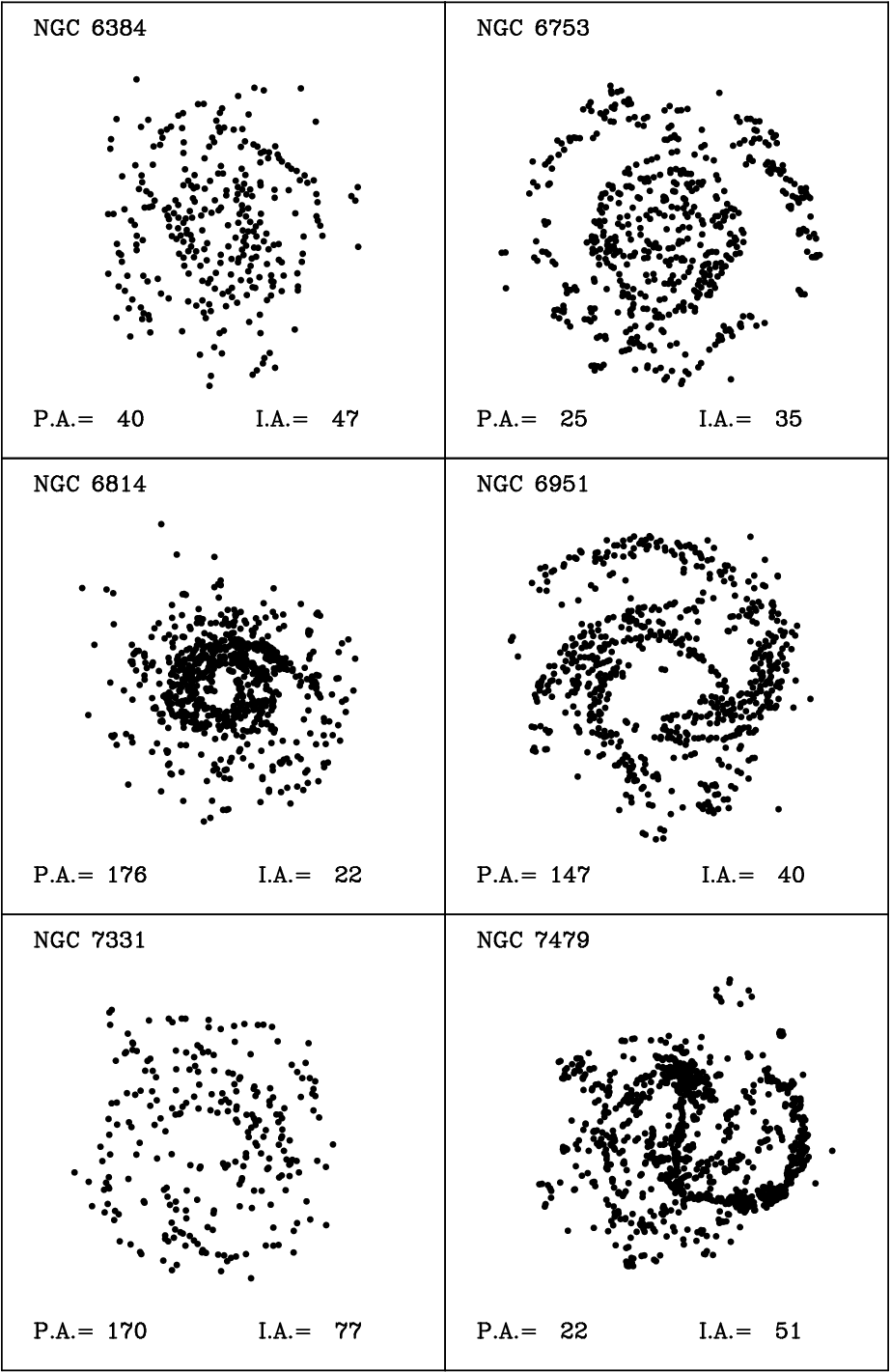}
\caption{Continued.}
\end{figure}

\begin{figure}
\includegraphics[scale=0.45]{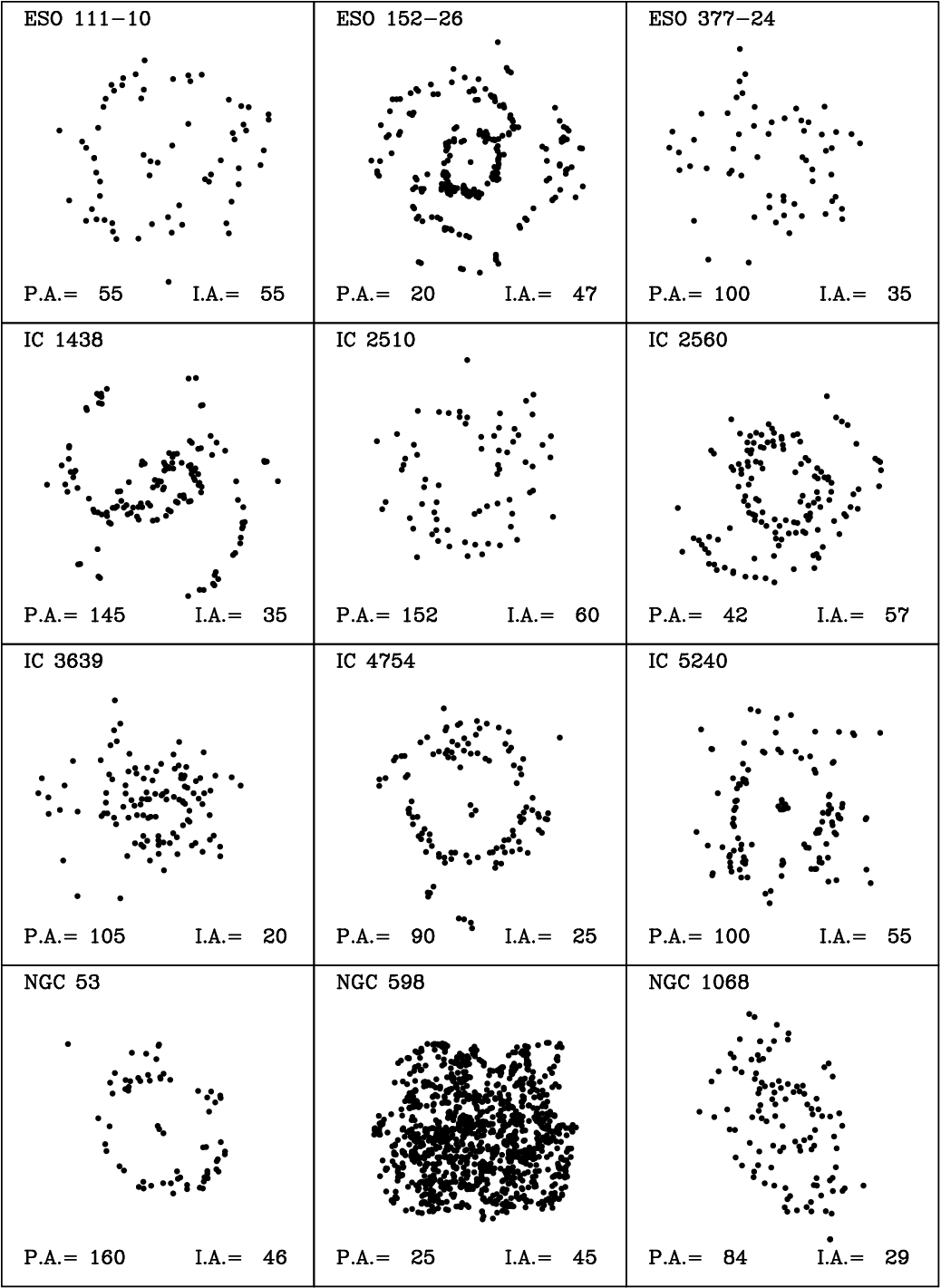}
\caption{Deprojected HII region distribution of the rest of the
galaxies in our sample.}
\label{album2}
\end{figure}

\begin{figure}
\addtocounter{figure}{-1}
\includegraphics[scale=0.45]{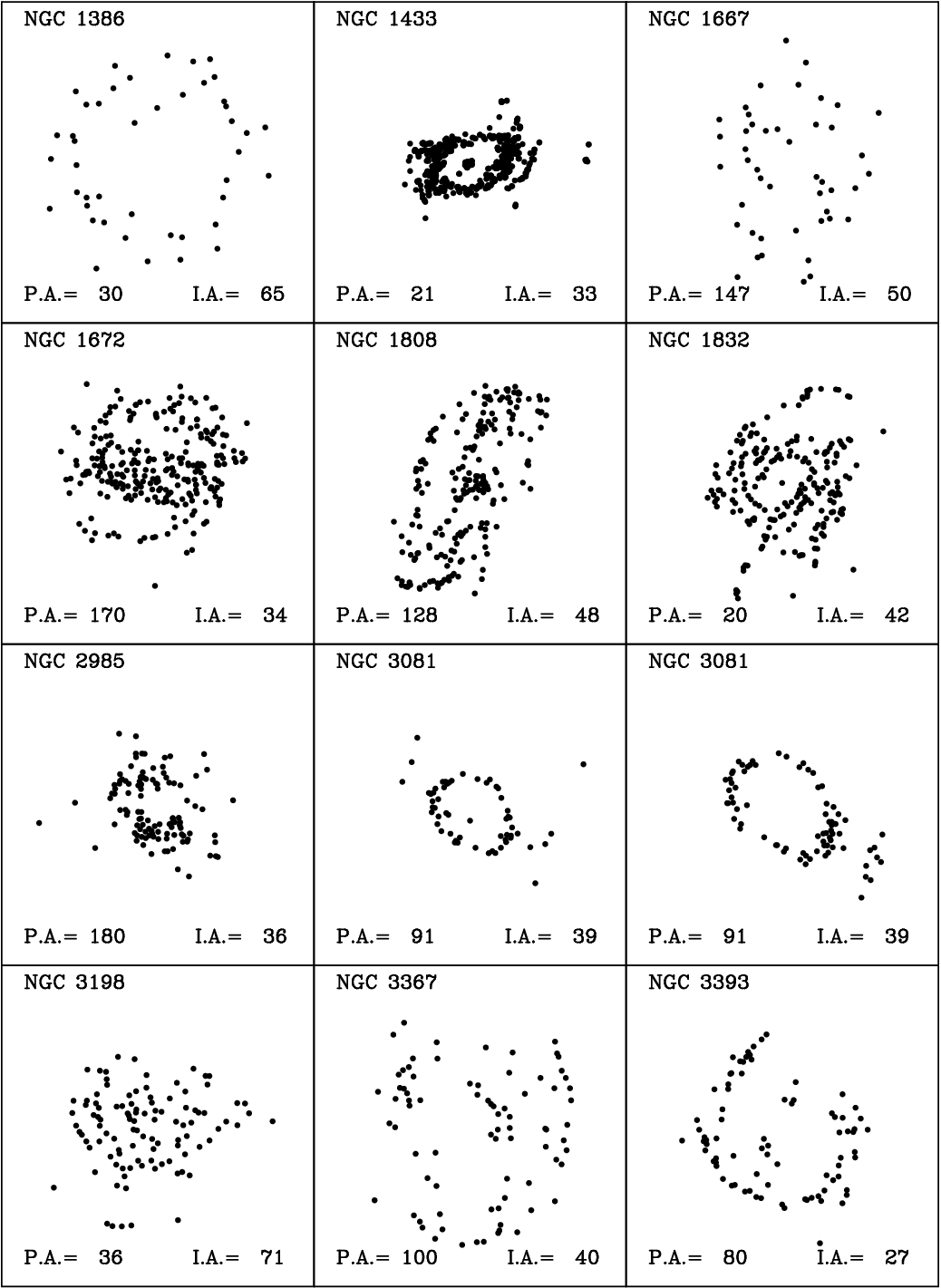}
\caption{Continued.}
\end{figure}

\begin{figure}
\addtocounter{figure}{-1}
\includegraphics[scale=0.45]{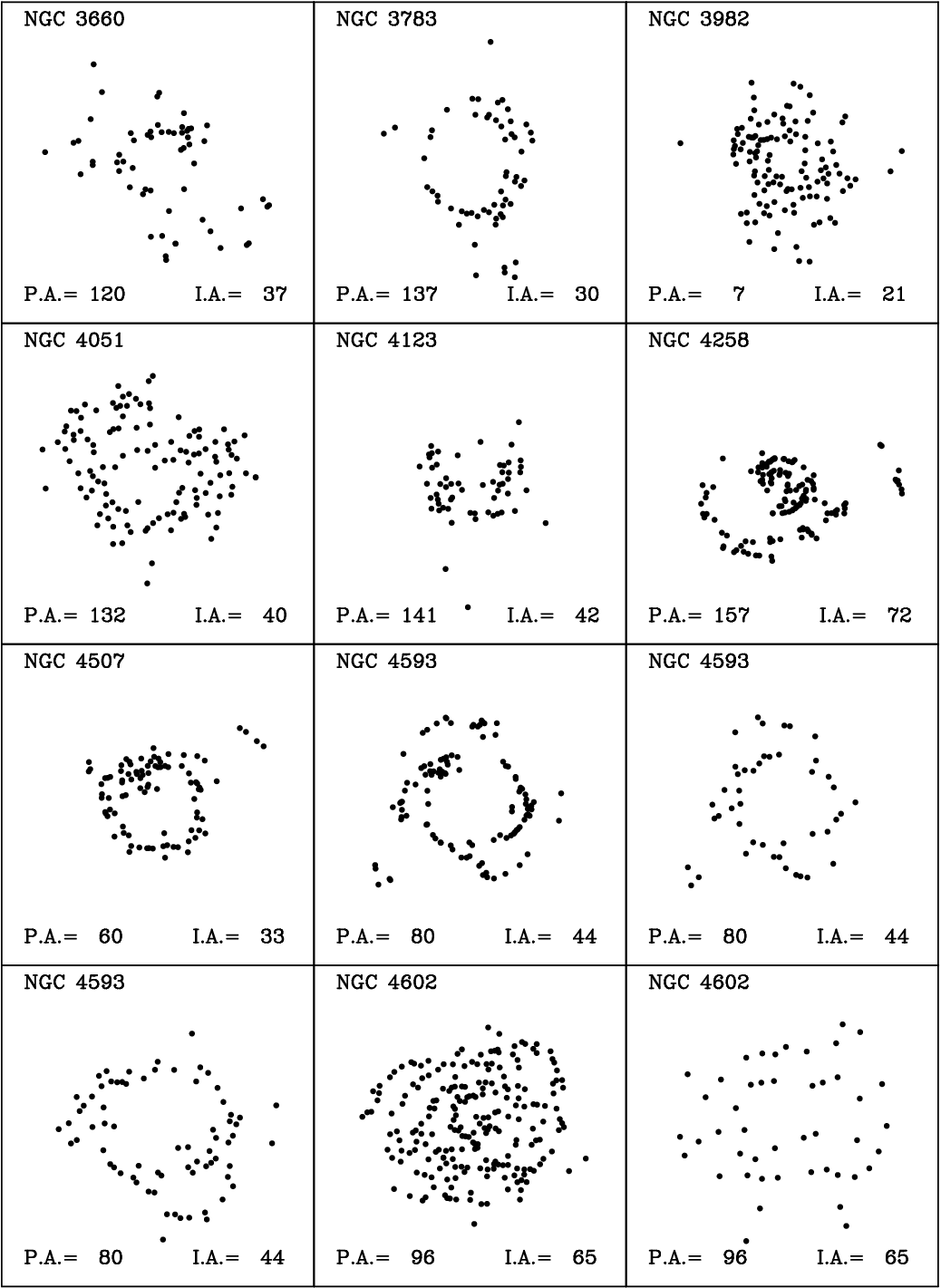}
\caption{Continued.}
\end{figure}

\begin{figure}
\addtocounter{figure}{-1}
\includegraphics[scale=0.45]{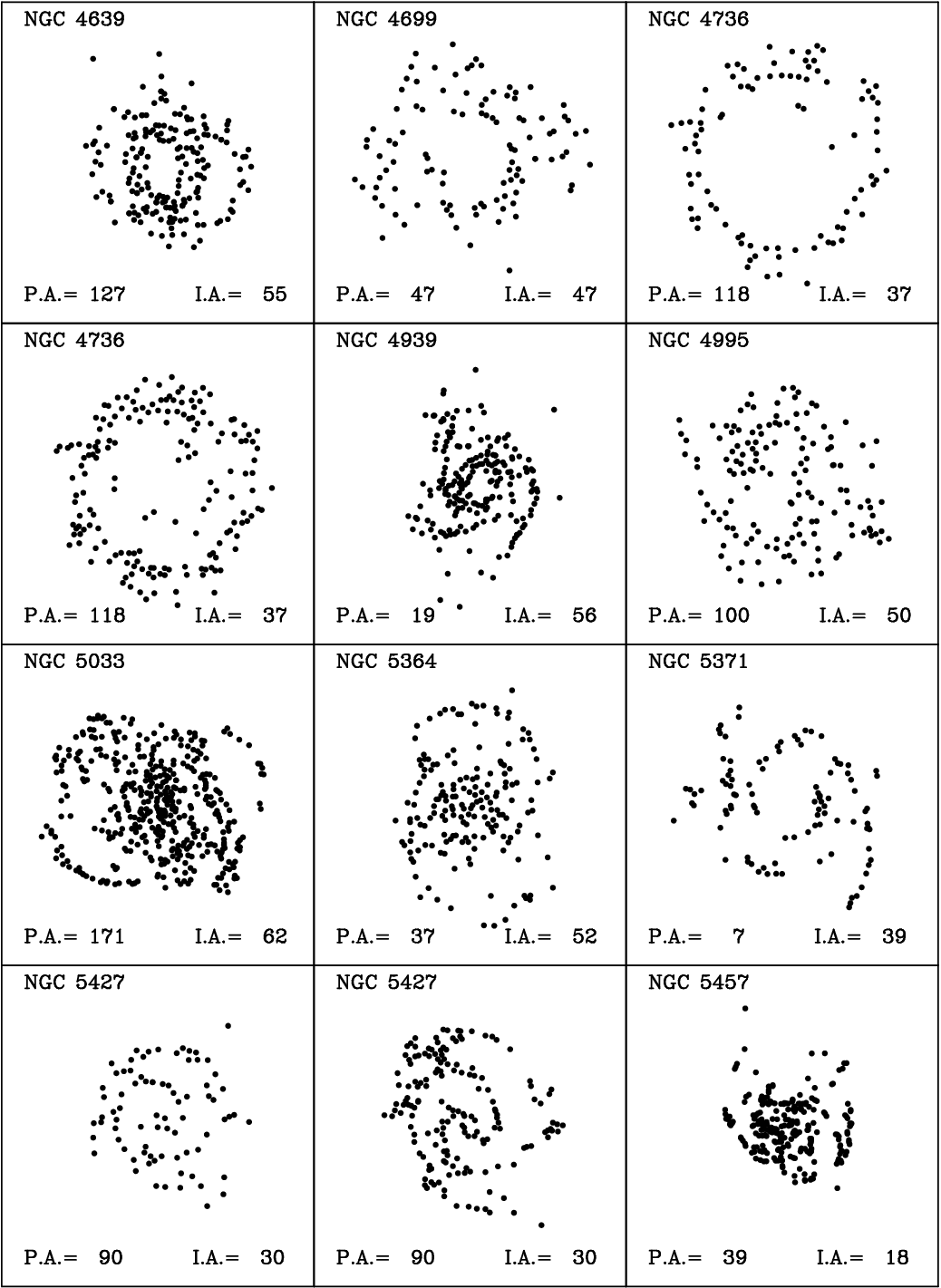}
\caption{Continued.}
\end{figure}

\begin{figure}
\addtocounter{figure}{-1}
\includegraphics[scale=0.45]{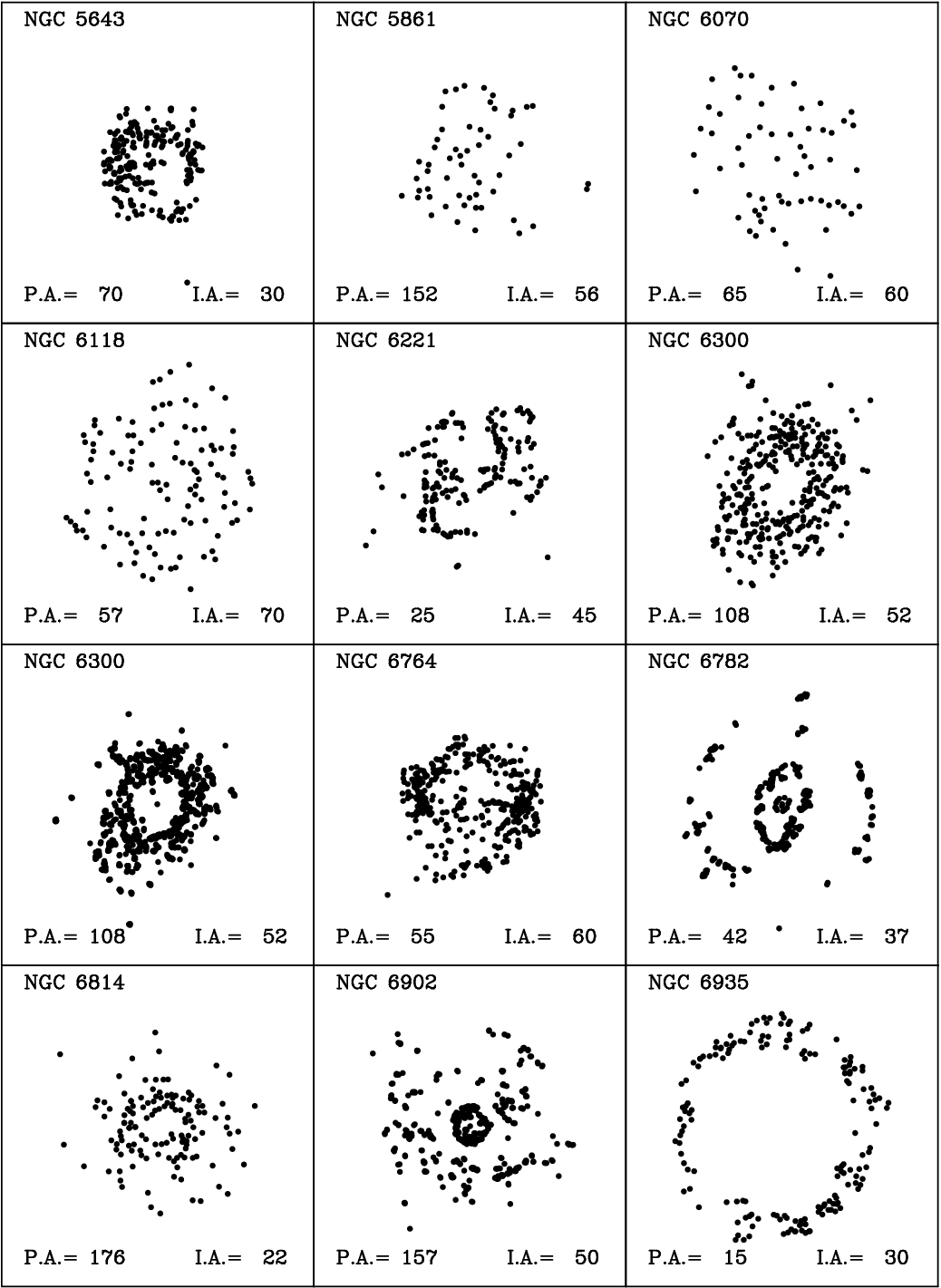}
\caption{Continued.}
\end{figure}

\begin{figure}
\addtocounter{figure}{-1}
\includegraphics[scale=0.45]{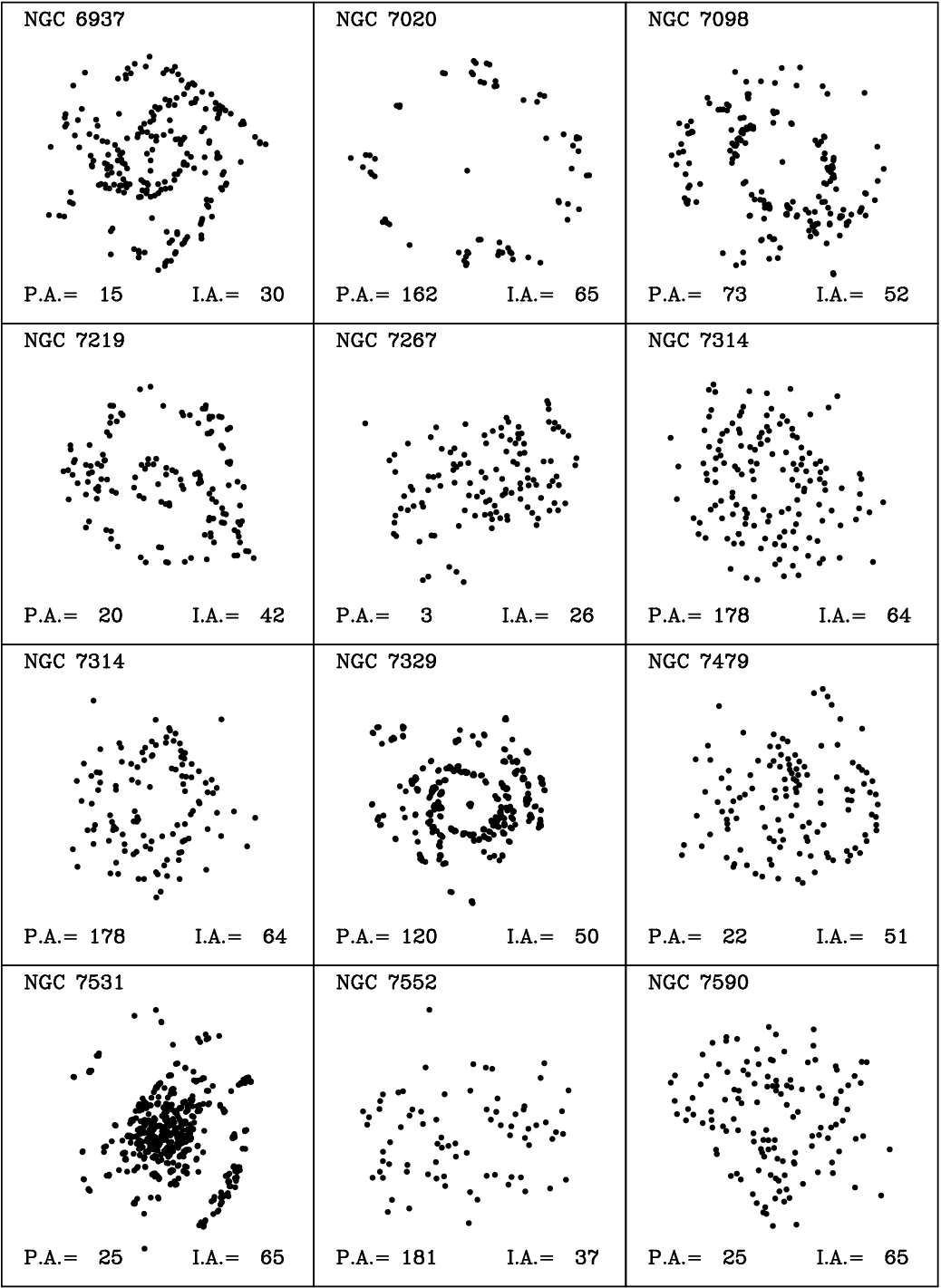}
\caption{Continued.}
\end{figure}

Searching in the literature, we can also find other kind of methods
to derive the deprojection angles that do not use the techniques described
above. Comte et al. (\cite{comte2}) used a
plot of the HII region distribution of M~101 in a $\log(r)-\theta$
plane to fit a straight line to the arms, using the hypothesis that
the arms are well described by logarithmic spirals. Iye et
al. (\cite{iye}) applied the two dimensional Fourier analysis
to the galaxy \object{NGC~4254}, using a photometric
image, and chose the deprojection angles that maximize the axisymmetric
component. Consid\`ere \& Athanassoula (\cite{considere1}) also applied 
Fourier analysis but used instead published catalogs of HII
regions. The criterion that they used was to maximize the signal to noise 
ratio in the $m=2$ component. This same criterion was used by
Consid\`ere \& Athanassoula (\cite{considere2}) but using instead photometric
galactic images.

In this paper we will use two methods, which were already introduced
by GGA (\cite{gga1}). For our first method we use two dimensional Fourier analysis
of the galaxy image. We decompose the HII region distribution in its
spiral components using a basis formed of logarithmic spirals. Since the
deprojected galaxy should be more axisymmetric than the projected one, we can
calculate the deprojection angles are as the angles that maximize the ratio:
\begin{equation}
\frac{B(0)}{\sum_{m=1}^{N} B(m)},
\end{equation}
where the $B(m), m=0,\ldots,N$ are defined as:
\begin{equation}
B(m) = \int_{-\infty}^{\infty} \mid A(p,m)\mid\,\, dp
\end{equation}
and the function $A(p,m)$ is the Fourier transform of the HII region
distribution, considered as a two dimensional distribution of
$\delta$-functions of the same weight:
\begin{eqnarray}
A(p,m) & = &\int_{-\infty}^{+\infty}\int_{-\pi}^{+\pi}\frac{1}{N}\nonumber\\
& \times & \sum_{j=1}^{N}\delta(u-u_j)\delta(\theta-\theta_j)e^{-i(pu+m\theta)}\,\,dud\theta \\
& = &\frac{1}{N}\sum_{j=1}^N e^{-i(pu_j+m\theta_j).}\nonumber
\end{eqnarray}

In the above $(r_j,\theta_j)$ are the polar coordinates of each HII region
on the galaxy plane, $u_j=\ln r_j$ and $N$ is the total number of HII
regions. The details of this method are outlined in GGA (\cite{gga1}). It is clear
that this method will work mainly for distributions which 
have a clear signal of all
the components of the spectrum, specially the $m=0$ component which
corresponds to the background disk distribution.

Our second method was also described in GGA (\cite{gga1}). This method is specially
devised for HII region distributions. If an axially symmetric
distribution of points is divided in $N_s$ equal sectors, like cake
pieces, and it is sufficiently rich, then we expect to have a
roughly equal number of HII regions in each piece, when viewing it
from a line of sight perpendicular to the disc. If the same
distribution of points is now viewed from a skew angle and we again use
sectors, which are of equal surface in the plane perpendicular to the
line of sight, then the number of HII regions will {\it not} be the
same in all sectors. Thus the correct deprojection angles will be those for
which the number of points in each sector is roughly the same, or, in
other words, when the dispersion around the mean value is minimum.

None of the methods described are free from systematic errors. 
In fact they all work perfectly well for the theoretical case of a razor 
thin, axisymmetric disc in circular motion around its center but present more or
less important problems for galaxies deviating from this idealized case.
When we fit ellipses to the isophotes, the
presence of strong bars and/or spiral arms can bias the results, if
the isophotes are not sufficiently external. However, Athanassoula and
Misiriotis (\cite{athanassoula2}) show 
that even for very strong bars the isophotes in the outermost part of
the disc are sufficiently circular to be used for deprojection. If we
use the velocity fields in HI, the external parts of the galaxy may be
warped, in which case the deprojection angles will be ill defined. Our
methods also can suffer from systematic problems. For instance, in the
case of the first method, if the HII region distribution delineates
mainly the $m=2$ arms and there are only few regions in the background
disk, the ratio (1) will not be well defined and the method can find
difficulties. Our methods can also find problems in the case of highly
irregular distributions, as well as in strongly barred
galaxies where many HII regions are concentrated in the bar
region. In these cases our methods will erroneously tend to circularize the bar.
A similar difficulty appears when many HII regions are located in
some galaxy rings. Buta (\cite{buta4}) using his Catalog of Southern Ringed 
Galaxies showed that rings need not be circular.
The outer features have an
average intrinsic axis ratio of $0.87\pm0.14$ while the inner features an
average of $0.84\pm0.10$. This indicates that some rings must have an oval
nature. Our methods, however, will tend to circularize these structures 
when they dominate the HII region distribution. Thus they will attribute to a
face on case an IA of $12.5$ degrees if there was an outer ring of axial ratio 
of $0.87$ and of $9.8$ degrees if there was an inner ring of axial ratio $0.84$.
Of course if only part of the HII regions are in the ring component, then the 
error could be smaller. In such cases, where HII regions delineate the several
components, we obtain secondary minima 
in the results obtained by our methods. It is thus necessary
to check if some of these minima give better results than the
principal minimum, which can be strongly biased by some of the effects
described above.

Thus, instead of relying on a single method in
the crucial step of deprojecting the galaxy image, a comparison of the
values given by the different methods is in order. In this way we can
choose the pair of deprojection angles that suit best a particular
galaxy. Our two methods, in conjunction with the literature values, were
used in our first sample of HII region distribution of spiral galaxies
(GGA, \cite{gga1}). In this
paper we will use this procedure to the HII region catalogs
published in the last decade. We should, however keep in mind that
our new sample has a high fraction of barred and ringed galaxies, 
where our two methods may not
perform as well as for the galaxies in the GGA (\cite{gga1}) sample 
(see Fig.~\ref{figbar}). 

\section{Notes on Individual Galaxies}

In this section we give some comments on the deprojection angles
chosen for each individual galaxy. This information is shown in
Table~\ref{Tangles}. In {\sl Column}~1 we give the galaxy name. In 
{\sl Columns}~2 
and~3 we show the PA and IA obtained using our two methods. In the first row we 
show the values for our first method, while in the second row those for our 
second method. If there are more than one catalog for a particular galaxy, the
values found using our methods are displayed in the following lines. 
In {\sl Columns}~4 and~5 we show the main values of PA and IA 
respectively found in the literature. In {\sl Column}~6 we give a key to
describe the method used to obtain the literature values. P is used for
photometric values, KH is used for values determined using HI 
velocity field, KC for values determined using a velocity
field in CO, KO for optical velocity fields, KS is used when the values come
from slit spectra determinations and finally O is used for
methods different to the previous ones. In {\sl Column}~7 we give a key for
the reference where this particular determination can be found. This key is
resolved in Table~\ref{Tref}. Finally in {\sl Columns}~8 and~9 we give the adopted
values of the PA and IA respectively. This same structure is repeated in the
second group of columns. Using the finally adopted values we can deproject the
catalogs of HII regions. The deprojected distribution of the richer 
catalogs showing with spiral structure are shown in Fig.~\ref{album1} while 
in Fig.~\ref{album2} we present the rest.
 
{\bf \object{ESO 111-10}}: This is a galaxy with apparent small size 
and the catalog of
HII regions is not very rich. Nevertheless, our two methods give a good
agreement when we use a secondary minimum for the second method. We adopt
the mean of both methods, which agrees well with RC3.

{\bf \object{ESO 152-26}}: For this galaxy the HII regions are placed mainly
in
the arms and in the inner ring. Our two methods give the same value for
the position angle and close values for the inclination angle. Eye estimates, 
however, show that the value of the PA is not satisfactory,
presumably because our methods try to circularize the ring. We thus use the
PA from the photometry of Crocker et al. (\cite{crocker}) and the mean
of our IAs, which is also the mean of the literature values.

{\bf \object{ESO 377-24}}: This is a small sized and not very inclined 
galaxy. The
number of HII regions in the catalog is also small and both methods
give a minimum at values more appropriate for a more inclined galaxy. Using
secondary minima for both methods we get a reasonable agreement with
the values from RC3 (\cite{devaucouleurs}). We finally adopt the values
from the second method, which are in agreement with the photometry and
close to the values of the first method.

{\bf \object{IC 1438}}: The HII regions are placed mainly in
the arms and the ring of this nearly face-on galaxy. The two
methods are in good agreement. But as there is no background disk
of HII regions, the first method may be biased by the presence of the
spiral structure. We thus keep the PA from the second method, which 
coincides with that found by Crocker et al. (\cite{crocker}). For the IA 
we take the average of our two methods.

{\bf \object{IC 2510}}: This is a galaxy of a very small apparent size.
Nevertheless, as it is quite inclined, our two methods give results in
good agreement and we keep the mean of the two, which is, furthermore,
in good agreement with the literature values.

{\bf \object{IC 2560}}: Our two methods are in good agreement for this
galaxy and we adopt the mean of their values, which agrees well with
the PA given in the RC3 (\cite{devaucouleurs}) and the Lauberts-ESO 
catalogs (\cite{lauberts}), and with the IA of the former.

{\bf \object{IC 3639}}: This is a nearly face-on galaxy in a small
group. The catalog of HII regions is quite irregular and does not
show any spiral structure. We adopt the values from our
second method, which are in agreement with the PA of Hunt et al. (\cite{hunt})
and the average IA of all methods. Note, however, that there is a lot of
dispersion around the mean values, which could mean that our estimate is not
very safe.

{\bf \object{IC 4754}}: This is a ringed, small size galaxy. Nearly all of the
HII regions of the catalog are placed in the external ring. Our methods
are in agreement for the IA and the agreement for the PA is only reasonable,
but the values from the PA are poorly determined as the projected catalog
looks quite round. We prefer to keep the values from the second
method, which are in agreement with two of the photometric values for the
PA and all the values for the IA.

{\bf \object{IC 5240}}: This case is quite similar to the previous
one. Again the HII regions are placed mainly in the external ring, but
as the galaxy is quite inclined, both methods are in reasonable agreement.
As before, we keep the values from the second method, whose PA is
in excellent agreement with all the literature values, and the IA with 
the results of the Lauberts ESO catalog (\cite{lauberts}).

{\bf \object{NGC 53}}: The HII regions are placed only in the outer ring and
their number is quite low. On the galaxy image, it seems that the inner
ring is not oriented as the main disk, but along the bar. Our first method
tends to circularize this ring which is not necessarily circular. So, we
keep the mean values from the
photometry of the RC3 (\cite{devaucouleurs}) and the Lauberts ESO 
catalog (\cite{lauberts}).

{\bf \object{NGC 157}}: This galaxy seems to be a bit irregular and
this is reflected in the rich catalog of HII regions. Nevertheless
the two methods are in excellent agreement between them and 
and in fair agreement with the literature values
Thus, we keep the mean of our values.

{\bf \object{NGC 210}}: The HII regions trace very well the arms and
the inner ring. Our two methods are in agreement with the literature
values and thus we keep the mean of our values.

{\bf \object{NGC 598}}: Our two methods are in good agreement but the values
that they give are not in agreement with the rest of the literature values,
probably due to an incomplete sampling of the galaxy disk in the HII region 
catalog from Hodge (\cite{hodge2}). Thus, we prefer to keep the mean of the
kinematically derived values.

{\bf \object{NGC 1068}}: The HII regions are placed only in the inner bright
oval, and the values obtained using both methods, while in agreement,
are inadequate. The values given by S\'anchez-Portal et al. (\cite{sanchez})
also pertain to the inner bright oval. Thus, we take the mean of the 
values from the velocity fields.

{\bf \object{NGC 1097}}: Our two methods agree well for the values of
IA, but the discrepancy is higher for the PA. The first method gives
rounder arms, so its PA and IA values must have been highly influenced 
by a Stocke's effect (Stocke \cite{stocke}), while
the second gives a rounder central part. These latter values agree 
very well with the HI kinematical values from Ondrechen (\cite{ondrechen})
as well as with three of the
photometrical estimates, so we will adopt the values from the second 
method.

{\bf \object{NGC 1386}}: Our two methods give identical results for this
quite inclined galaxy, which are furthermore in agreement with the 
photometric estimates. We adopt the values from our two methods. 

{\bf \object{NGC 1433}}: The HII regions trace very well the inner ring of
this galaxy. Both methods try to circularize this inner ring giving a strong
disagreement with the kinematical values. We adopt the values from
the velocity fields.

{\bf \object{NGC 1566}}: This galaxy was also studied in the first
paper, where we used the catalog by Comte et al. (\cite{comte2})
which traces also the external arms. For this new catalog, the HII
regions are placed mainly in the inner oval part and the results of
our methods
disagree. The values obtained from the first catalog by GGA (\cite{gga1})
are in good agreement with the values from the kinematics, so we adopt the 
mean values of the GGA values and the optical velocity field from 
Pence et al. (\cite{pence2}).

{\bf \object{NGC 1667}}: This galaxy has a small apparent size, with a
catalog with a low number of HII regions. Nevertheless, as the
galaxy is quite inclined, the two methods are in reasonable agreement and
we give the mean of both methods.

{\bf \object{NGC 1672}}:  Our second method does not converge, probably due
to the strong bar present in this galaxy. On the other hand our first 
method gives a PA value in good agreement with the photometric values form
RC3 (\cite{devaucouleurs}) and the Lauberts-ESO catalogs (\cite{lauberts}),
but the value obtained for the IA value seems too high. We keep the values
from the RC3 catalog.

{\bf \object{NGC 1808}}: There is a good agreement between the two methods, 
but the HII regions trace only the inner bright oval part and
there are no regions in the outer arms or disk. The outer disk in the galaxy image
seems to be much less inclined. For this reason we prefer to keep the
kinematic values from Koribalski et al. (\cite{koribalski}).

{\bf \object{NGC 1832}}: This galaxy was studied also in the first paper, but
using a catalog with a low number of HII regions. Using this new
catalog, the two methods are in good agreement, and there is a rough
general agreement with the rest of the values from the literature. Thus
we keep the mean of our two methods.

{\bf \object{NGC 2985}}: The HII region distribution is quite irregular, but
despite this fact, the first method gives values in reasonable 
agreement with the
RC3 (\cite{devaucouleurs}) catalog. The second method does not work.
We adopt the mean between our first method and RC3 (\cite{devaucouleurs}).

{\bf \object{NGC 2997}}: This galaxy was also studied in the first
paper using a similar catalog. Using the new catalog we find values
for the IA that are in reasonable agreement between them, while the values 
for the PA
are not so well constrained. The mean of the two methods are in
agreement with the kinematical values from Milliard \& Marcelin
(\cite{milliard}) and the photometric values from
the RC3 (\cite{devaucouleurs}) and, to a lesser extent, with the
rest of the values. We adopt the mean of our two methods.

{\bf \object{NGC 3081}}: For this galaxy we have two catalogs with
distributions of similar shape, the HII regions being mainly in the
inner ring. As both methods try to circularize this ring we prefer to
adopt the mean of the values from the photometry and kinematics from
Buta \& Purcell (\cite{buta5}). 

{\bf \object{NGC 3198}}: There is a good agreement between our two methods
for this  quite inclined galaxy. There is also a general agreement
with the rest of methods. The projected HII region distribution,
however, seems somewhat irregular, and this may bias our two methods,
giving higher values from the PA. For this reason we prefer to adopt
the mean values from the HI kinematical studies of Bosma (\cite{bosma1}) 
and Begeman (\cite{begeman1}). Note that these are in good agreement
with our second method and reasonable agreement with the
first.

{\bf \object{NGC 3359}}: Our methods are not adequate for this galaxy
because a lot of HII regions are located in the bar region. We thus
keep the mean of the kinematical values from the H$\alpha$ velocity field
from Rozas et al. (\cite{rozas3}) and HI velocity fields from Gottesman 
(\cite{gottesman}) and Ball (\cite{ball}). 

{\bf \object{NGC 3367}}: The values given by the various methods cover a broad
range of values.
Our two methods give the same value for the IA,
which is furthermore in agreement with
the value from Danver (\cite{danver}). The average of the two
values of the PA are in agreement with the  photometry
from Grosb\o l (\cite{grosbol1}). We thus adopt the mean of our two methods.

{\bf \object{NGC 3393}}: The situation for this galaxy is highly
unsatisfactory. The PA of 160 is mainly reflecting the orientation of the
bar, so it can be dismissed. The estimate of $41$ degrees comes from the
outer isophote, but this seems to be heavily distorted, partly by the arms. 
Our methods should suffer from Stocke's effect (\cite{stocke}). Since our 
purpose in the following pages will mainly be to study the spiral structure
and our two methods agree well, we will adopt their average for our future
work. We stress, however, that this is due to the lack of any better
estimate.

{\bf \object{NGC 3631}}: This is a nearly face-on galaxy, and thus although
we use a very rich catalog, the values of PA and IA are not very well
constrained. We adopt the value of the PA derived in the kinematical analysis
of Knapen (\cite{knapen1}). However, the kinematic analysis did not
give a value for the IA, so we adopt the mean of our methods for this angle,
which is in rough average with the photometric values.

{\bf \object{NGC 3660}}:  The catalog of HII regions is quite poor and the
values of PA and IA are not very well constrained. Nevertheless, the two
methods are in good agreement and in rough agreement with the photometric
values from RC3 (\cite{devaucouleurs}). We adopt the mean of our two methods.

{\bf \object{NGC 3783}}: The HII regions of this catalog populate an
inner ring. Nevertheless, our two methods are in good agreement with
the only PA value that we found in the literature and with the IA of 
RC3 (\cite{devaucouleurs}). There is also a rough
agreement with the IA values from other studies. We adopt
the mean of our methods.

{\bf \object{NGC 3982}}: This is a small apparent size galaxy nearly
face-on and the
HII region catalog is quite poor. Both our methods have two clear minima,
however, and are in good agreement between them. Moreover, they are in 
rough agreement with
the photometry, so we adopt the mean values of all methods.

{\bf \object{NGC 4051}}: The HII region distribution is somewhat
irregular and the first method gives results only in rough agreement
with the values from the second method. We finally adopt the kinematical
values of Listz \& Dickey (\cite{listz}), which are in general agreement
with most of the photometric values.

{\bf \object{NGC 4123}}: The number of HII regions is quite low and
the HII region distribution traces mainly the bar region. The first
method does not work properly, so should be neglected. The values from
our second method are in good agreement with the values from the kinematics.
Seen the poor quality of the HII region catalog
we adopt the mean values from the kinematics.

{\bf \object{NGC 4258}}: As our two methods are in agreement for this
quite inclined galaxy we adopt their mean value which is in agreement 
with the results of the velocity fields of van Albada (\cite{vanalbada1}) 
and van Albada \& Shane (\cite{vanalbada2}).

{\bf \object{NGC 4321 (M100)}}: This galaxy is not very inclined and 
thus the value of the PA is not well constrained. On the other hand,
there is a rough agreement about the IA for the rest of the methods.
We adopt the kinematical values of Guhatakurta (\cite{guhatakurta}).

{\bf \object{NGC 4507}}: The galaxy has a small size and the HII regions
populate a ring. Our two methods do not work properly because they try to
circularize the ring. We adopt the deprojection angles obtained in the
photometric study of Schmitt \& Kinney (\cite{schmitt}).

{\bf \object{NGC 4593}}: This is a strongly barred galaxy for which there are
three published catalogs of HII regions. Our two methods are in good
agreement for the richer catalogs but the first method gives discordant
results in the second catalog and the second method does not work at all.
Thus we discard the values from the second catalog. The mean of our values
are also in general agreement with the photometric values, except for the
case of the I photometry by Schmitt \& Kinney (\cite{schmitt}) who give
a higher PA value than the rest. We adopt the mean values of the results 
of our two methods applied to the first and third catalog.

{\bf \object{NGC 4602}}: All values of the PA are in rough agreement,
except for the value given by Mathewson \& Ford (\cite{mathewson}),
which by eyeball estimate does not look very reasonable. We
adopt an average value of our two methods and both catalogs, which
also agrees with RC3 (\cite{devaucouleurs}) and Danver
(\cite{danver}).

{\bf \object{NGC 4639}}: There is good agreement between the values 
obtained by our methods. Thus, we take the average values. This is also
in agreement with the literature values.

{\bf \object{NGC 4699}}: We adopt the average of our two methods, which is in
good agreement with the photometric values.

{\bf \object{NGC 4736}}: For this galaxy, we have two catalogs. In both
cases, nearly all the HII regions are placed in the external ring.  
Our four results are only in rough agreement and they are, furthermore, 
not very reliable since they pertain to the ring.
Thus, we prefer to adopt, the mean values
given from the velocity fields (Bosma et al. \cite{bosma3},
Mulder \& van Driel \cite{mulder}, Buta \cite{buta3}) which are in
good agreement between them and also with the average of our two
methods and the two catalogs.

{\bf \object{NGC 4939}}: We have two catalogs of similar richness for this
galaxy. It is quite inclined and the two methods are in rough
agreement for both catalogs. We take the averages of our two methods
and the two catalogs, which is in agreement with the
photometric values and the value given in GGA (\cite{gga1}).

{\bf \object{NGC 4995}}: Our two methods give identical results, which we
adopt, since they are also in good agreement with the photometric
values from RC3 (\cite{devaucouleurs}) and Grosb\o l (\cite{grosbol1}).

{\bf \object{NGC 5033}}: A number of estimates are available for the
deprojection angles of this galaxy, and all cluster in a relatively
narrow range of values. The results of our two methods are in good 
agreement between them and with the rest of the estimates. The deprojection
of the HII region distribution is particularly sensitive to the adopted
value of the IA. We tired several averages of the individual estimates both
straight and weighted by our judgment of the quality which resulted in 
identical values. We adopt these mean values. 

{\bf \object{NGC 5194 (M51)}}: Our two methods are in good agreement, but
the values of the position angles in particular, using the primary minima,
are quite discordant with the values obtained from the analysis of the
velocity fields. Note that they are also similar to some of the derived 
photometric values. The deprojected galaxy using our values looks quite
good, as already noted in GGA (\cite{gga1}). There is, however, a
secondary minimum, which is in agreement with the values derived from
the velocity fields. With these values we get also a round 
deprojected galaxy. We adopt the value from Rots et
al. (\cite{rots}), which are as stated, in agreement with the average
of the secondary minima value.

{\bf \object{NGC 5364}}: We take the average values from our two
methods, which is in agreement with the rest of the values in the
literature. 

{\bf \object{NGC 5371}}: The galaxy was too big to fit in the CCD
frame used by Gonz\`alez-Delgado et al. (\cite{gonzalez}), so that,
the PA values found using this catalog can not be very reliable. We
thus adopt for this galaxy an average of the photometric values.

{\bf \object{NGC 5427}}: Here the two methods and the three catalogs give
different results, which again are very different from the literature
values. This is not surprising as the galaxy is not very inclined and
there is no derived velocity field in the literature. For lack of
any stronger criteria and looking at the deprojected images obtained
with all these values, we decided to adopt the result of the first
method and the richer catalog. However, it should be stressed that
these are very ill defined values.

{\bf \object{NGC 5457 (M101)}}:  This galaxy is nearly face-on. Thus, 
the values of the PA are not well constrained. Our methods prefer a galaxy
completely face on but, as the galaxy is quite asymmetrical, 
we have decided to adopt the values obtained by the HI high resolution 
study of Bosma et al. (\cite{bosma2}). 

{\bf \object{NGC 5643}}: The catalog of HII regions has an irregular
distribution. The outer parts look also irregular in the galaxy
images. Thus, our two methods as well as the photometric values are
very unsafe. For lack of any stronger criteria, we adopt the results of 
the first method, which give a rounder deprojected object. It should
nevertheless stressed that is a very unsafe estimate.

{\bf \object{NGC 5861}}: There are very few HII regions in this
catalog, so we adopt the values given by Grosb\o l \& Patsis
(\cite{grosbol2}). It should, nevertheless be noted that, although
the catalog is poor, our results are in fair agreement with the
photometric ones.

{\bf \object{NGC 6070}}: There is a good agreement between our two
methods and the results available in the literature. We adopt the
average of our two methods.

{\bf \object{NGC 6118}}: For this galaxy we have, as in the former
case, a good agreement between our methods and the literature
values. Thus, we adopt the average of our two methods.

{\bf \object{NGC 6221}}: The HII regions trace mainly the arms and
give a rather irregular distribution, so that our second method is not
reliable. There is a good agreement between our first method and the
photometric value from Pence \& Blackman (\cite{pence1}).  We finally
adopt our first method values. Note that the IA is in good agreement
with the remaining photometric values.

{\bf \object{NGC 6300}}: The results of the kinematical values of Ryder et 
al. (\cite{ryder1}) and of Buta (\cite{buta2}) are in perfect
agreement, so we adopt these values. Our two methods are less
reliable, since the HII regions are mainly lying in a
ring for both catalogs. Nevertheless, it is worth noting that there 
is a reasonable agreement between our values and the adopted ones and
also with the rest of the values.

{\bf \object{NGC 6384}}: Our two methods agree between them and we have
adopted the average of their values. Our IA values agree well with the
photometric ones, while the PA does so reasonably well. The IA given by 
Prieto et al. (\cite{prieto}) seems more reliable than that of 
S\'anchez-Portal et al. (\cite{sanchez}), since the data extend further 
out in the region which is less influenced by the oval bar. This agrees
with the mean of our values. The PA we adopt is somewhat 
higher than those given by the photometry. The deprojected catalog,
however, has a round shape and thus we keep our values.

{\bf \object{NGC 6753}}: Our two methods agree reasonably well between
them, so we take the average values which are in good agreement with
the photometric values.

{\bf \object{NGC 6764}}: Our two methods give identical results and
also agree quite well with the available photometric values, so we
adopt our values. 

{\bf \object{NGC 6782}}: We adopt the average of our two methods,
which is in reasonable agreement with the photometric values, as in
the previous galaxy.

{\bf \object{NGC 6814}}: This is a nearly face-on galaxy and as a result 
the values of the angles are not well constrained. We have two catalogs
of different richness. As the PA can not be well constrained using our
methods, we have decided to adopt the values from the kinematics of
Listz \& Dickey (\cite{listz}) which are in fair agreement with the values
obtained by our methods for the less rich catalog from Evans et al.
(\cite{evans}). Note that the richer catalog of Knapen et
al. (\cite{knapen3}) has a well delineated strong northern arm which
can bias our results.

{\bf \object{NGC 6902}}: We adopt the average of our two methods which 
is in good agreement with the results given in the RC3
(\cite{devaucouleurs}) catalog, and agrees also
with the average of the rest of the literature determinations.

{\bf \object{NGC 6935}}: The HII regions trace the outer ring
only. Our second method does not work, and converges to a value with
too high inclination, so we neglect it. We adopt the result from the
first method which is in good agreement with the inclination of the RC3
(\cite{devaucouleurs}) catalog.

{\bf \object{NGC 6937}}: No reasonable minimum was found by the first 
method and we adopt the values from the second method, which agree well 
with the inclination of the RC3 catalog, but not so well with the values 
quoted by Crocker et al. (\cite{crocker}).

{\bf \object{NGC 6951}}: We take the average of our two values which are 
in good agreement with the values derived by Grosb\o l (\cite{grosbol1}).

{\bf \object{NGC 7020}}: The few available estimates agree well with 
our two methods, so we take the average of our two values.

{\bf \object{NGC 7098}}: Our two methods try to circularize the inner 
ring so that, even though they agree well between them, they are not 
very meaningful. We thus adopt the average of the photometric results.

{\bf \object{NGC 7219}}: The first method does not work very well, so 
we list a secondary minimum. Our adopted values are the average of 
the two methods and are in reasonable agreement with the few available
literature values.

{\bf \object{NGC 7267}}: The few HII regions are mainly in and around 
the bar and not in the surrounding disc. We thus adopt, the value given by
Crocker et al. (\cite{crocker}), which is in reasonable agreement with the
other photometrically determined estimates.

{\bf \object{NGC 7314}}: Our two methods applied to the two catalogs give
results in very good agreement. We thus adopt their mean value, which
is also in good agreement with the few results available in the
literature.

{\bf \object{NGC 7329}}: Our second method converges to an inclination angle
which seems too large. We thus neglect it and take the result of the
first method, which is in reasonable agreement with the little that is
available in the literature.

{\bf \object{NGC 7331}}: The results of our two methods agree very well 
between
them, and with the main kinematic estimates. We thus adopt the average
of our two methods.

{\bf \object{NGC 7479}}: This is a difficult galaxy since, on top of a bar,
there is a major $m=1$ asymmetry, clearly seen in the deprojected image for
both catalogs. This can bias our methods, thus we prefer to
adopt the kinematical value
from Laine \& Gottesman (\cite{laine}), which is in good agreement with the
results given by our second method.

{\bf \object{NGC 7531}}: This galaxy is quite inclined, so it is crucial 
for the PA to be accurate. Our first method gives a value which is, by
visual inspection, not acceptable. We thus adopt the result of our
second method, which is in good agreement with the kinematic
values.

{\bf \object{NGC 7552}}: The HII regions are mainly concentrated in 
and around
the bar region and thus can not give useful information for the
deprojection. We thus adopt the values from RC3 (\cite{devaucouleurs}).

{\bf \object{NGC 7590}}: Our two methods give results that coincide, so we
adopt their average values, as they are also in reasonable agreement
with the few available photometric estimates.

\section{Discussion}

In this section we compare the different methods used to determine the
deprojection angles (PA, IA). Our aim is to see whether any one is
inferior or superior to the others. 
We compared 7 methods, or groups of methods.
Our two methods (method 1 and 2) constitute a group each. The values
obtained by Danver (\cite{danver}) form group D, those by Grosb\o l
(\cite{grosbol1}) form group G, and the values from the RC3 catalog
form group RC3. The sixth group 
(kinematics, K) includes all the values obtained using information from 
HI or optical velocity fields. Finally the seventh group (photometry,
P) includes all results obtained fitting ellipses to the outer parts
of the galaxies. The values in these last two groups do 
not  constitute homogeneous samples, but the methods used by the different 
authors are very similar. Moreover, the kinematical method is quite reliable, 
specially for the determination of the PA.

\begin{figure}[h]
\includegraphics[scale=0.5]{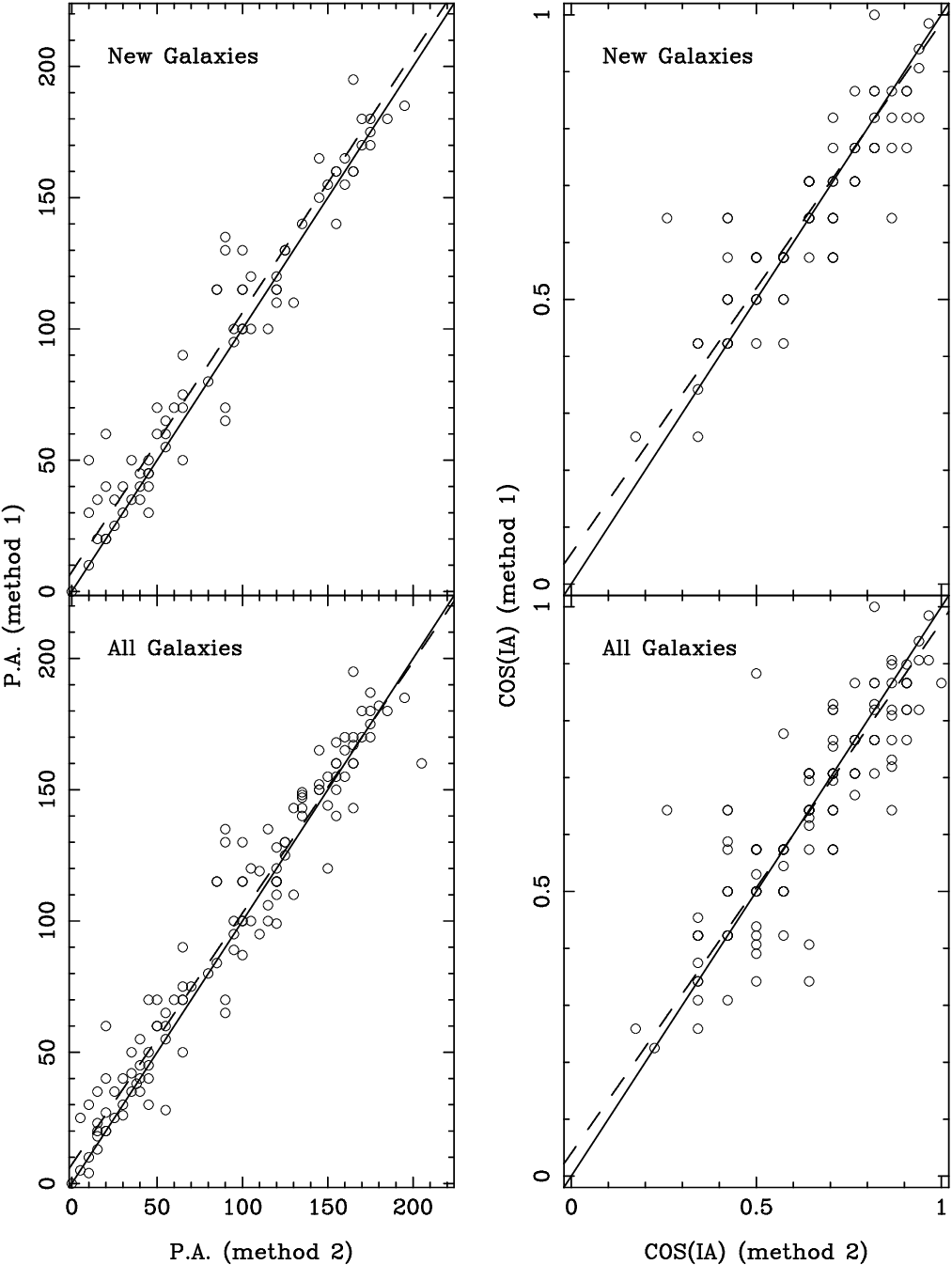}
\caption{Correlations between the values of the PA and IA derived
using our two methods. Solid line. Diagonal (perfect
correlation). Dashed line. Best fitting weighted linear correlation.}
\label{figours}
\end{figure}

\begin{figure}[th]
\includegraphics[scale=0.5]{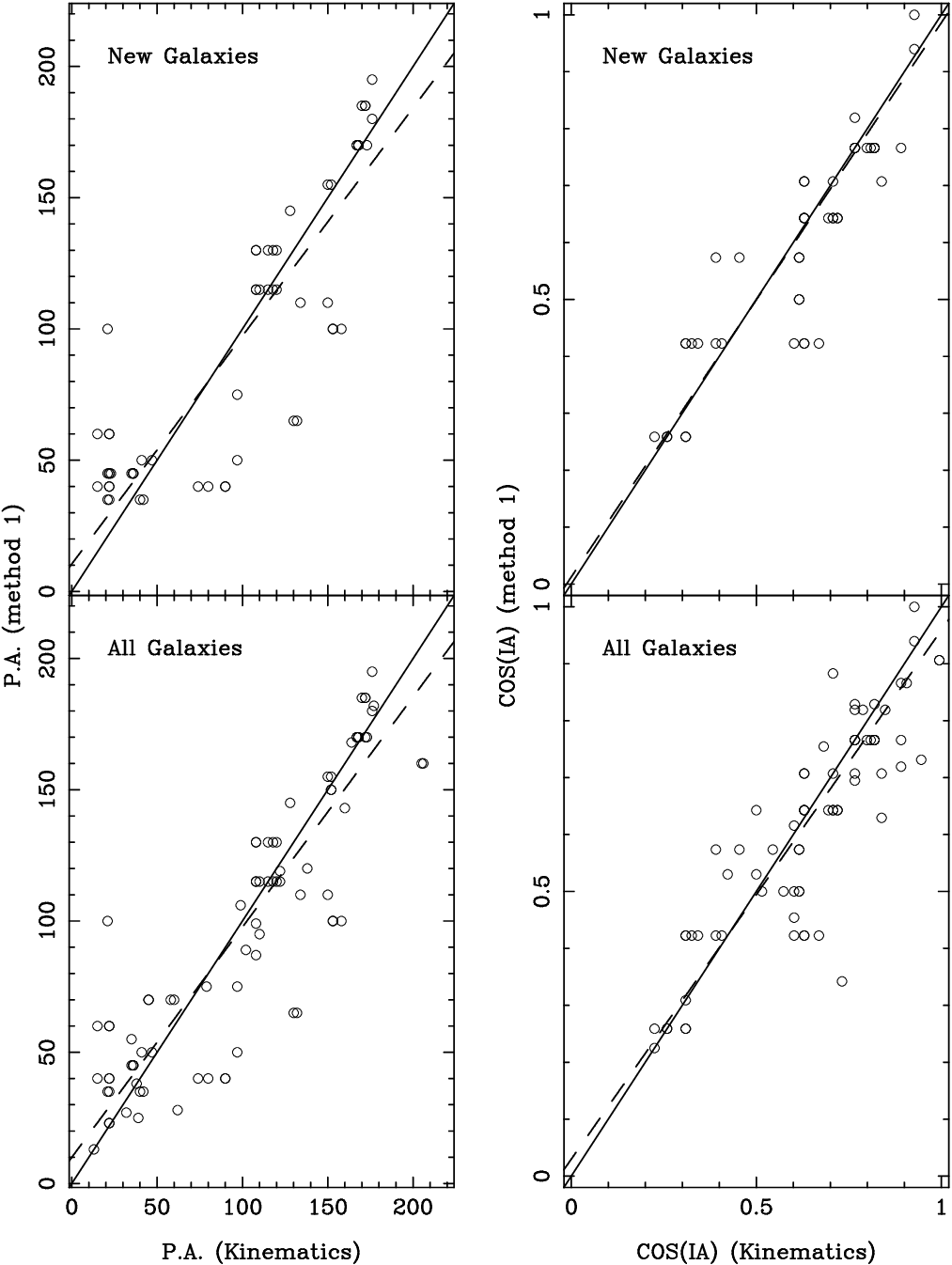}
\caption{Same as Fig~\ref{figours} but for the correlations between our
first method and the results from HI kinematics.}
\label{figok}
\end{figure}

\begin{figure}[h]
\includegraphics[scale=0.5]{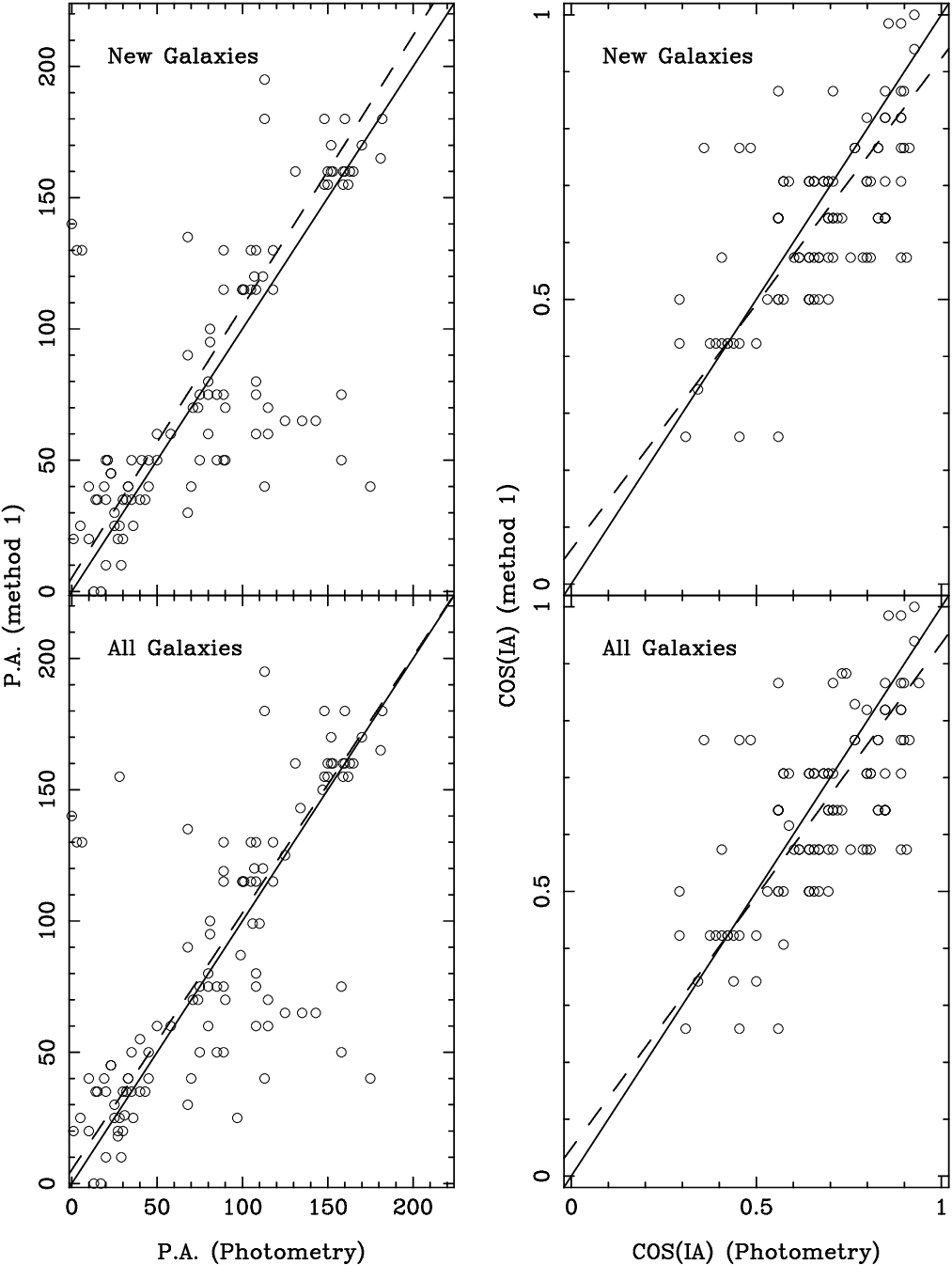}
\caption{Same as Fig~\ref{figop} but for the correlations between our
first method and the results from the photometry.}
\label{figop}
\end{figure}

For comparing any two methods we fitted a straight line to all pairs
of values, using a maximum 
likelihood algorithm which minimizes the $\chi^2$ merit function 
\begin{equation}
\chi^2(a,b)=\sum_{i=1}^N 
\frac{(y_i-a-bx_i)^2}{\sigma^2_{iy}+b^2\sigma^2_{xi}},
\end{equation}
where $\sigma^2_{xi}$ and $\sigma^2_{yi}$ are the errors for 
the $i$th value (Press et al. \cite{numerical}).
 
When using our values, we assigned a weight to each
catalog of HII regions as follows: If the
catalog is very irregular with a small number of HII regions we assigned
a weight of $0.1$. If the number of HII regions is small but the
catalog looked 
quite regular we assigned a weight of $0.3$. If the catalog has a fair number
of HII regions, but the distribution has some irregularities, we assigned
a weight of $0.6$ and, finally, if the catalog looked regular and had a 
fair number of HII regions we assigned a value of $1.0$. 
We introduced a further weight, this time for the PA values of all the
methods, to take into account the 
fact that for galaxies nearly face on, it is 
difficult for any of the methods to assign a reliable value of the
PA. We thus assigned a low weight to the PA values of nearly face-on
galaxies and a weight of 1. for all the rest. The errors were taken as
the inverse of the weights, or, in cases with two weights, as the inverse
of their product. 
In all the correlations we discarded the galaxy NGC~5194. For this
galaxy, the values determined for the PA and IA using the kinematical
information are in clear disagreement with the values determined using any of
the other methods. This must be  due to the strong interaction with
the companion, NGC~5195. 

Some illustrative correlations between pairs of methods are shown in Figures
\ref{figours} to \ref{figop}. In Fig~\ref{figours} we compare the
results of our two methods, in Fig~\ref{figok} our first method to the 
kinematical  values and in Fig~\ref{figop} our first method to the
photometrical determined values. 
To compare quantitatively the results of these and all the remaining
correlations (not shown here) we used
correlation coefficients as well as the
weighted mean of the orthogonal distances of all the points to the best 
fitting straight line. 
The results of the comparison are shown in Tables~\ref{distab} and
\ref{cortab}. Tables~\ref{atab} and \ref{btab} give the coefficients $a$ and
$b$ of the linear regressions. In all cases the values above and to
the right of 
the main diagonal correspond to the PA, and the values under and to
the left of the main diagonal correspond to the $cos($IA$)$. We chose
the $cos$, rather than the angle, since the former is uniformly 
distributed. Note that the values of the orthogonal distances and of
the coefficient $a$ for
the case of $cos($IA$)$ have been rescaled so as to make them
directly comparable to the corresponding values 
of the PA. For the comparisons we pooled together the data 
of the galaxies of the new sample presented in this paper and the galaxies of
the previous sample (GGA, \cite{gga1}). Obviously a perfect fit
between two methods would imply a correlation coefficient of 1, a
weighted mean of the orthogonal distances of zero, as well $a$ = 0. and
$b$ = 1. Random errors will introduce a scatter, which would lower the
correlation coefficient and raise the value of the weighted mean of
the orthogonal distances. Systematic differences, as would occur
e.g. if a given method systematically overestimated or underestimated
the geometric angles, would change the values of the coefficients $a$
and $b$.

\begin{center}
\setcounter{table}{3}
\begin{table}
\caption{ Weighted mean orthogonal distances for the PA correlations 
(upper triangle) and $cos($IA$)$ (lower triangle).} 
\label{distab}
\begin{tabular}{rrrrrrrr}\\
\hline
      & 1 st   &   2nd  &   RC3 & G     & D      &    K   & P  \\\hline
 1st  &        & $7.1$ &$10.5$&$11.3$& $9.6$ & $13.3$& $16.5$ \\
 2nd  & $7.4$ &        & $9.1$& $8.9$& $6.2$ & $11.4$& $14.1$ \\
 RC3  & $9.9$ &$10.1$ &       & $6.2$& $5.7$ &  $9.3$& $10.1$ \\
 G    & $9.2$ & $9.0$ & $6.1$&       &$11.6$ & $16.0$& $17.8$ \\
 D    &$13.7$ &$11.9$ & $5.9$&$12.1$&        &  $9.4$& $20.7$ \\
 K    & $9.5$ & $9.7$ & $5.0$& $6.8$& $9.7$ &        & $15.8$ \\
 P    &$12.9$ &$16.2$ & $7.7$&$10.1$&$11.3$ & $10.9$&     \\\hline
\end{tabular}
\end{table}
\end{center}

We note that, although no two methods we have used agree perfectly,
there are also no glaring discrepancies. No method seems to
overestimate or underestimate either of the deprojection angles. 
Also no method seems to give systematically smaller correlation
coefficients, which would be indicative of a large random error in the
results of this method. All correlations range between acceptable and
good. The worst correlation coefficient, 0.51, is found when comparing
the $cos$(IA) of Danver and Grosb\o l. This, however, is not due to one
or several discordant galaxies, but rather to the fact that the
galaxies are relatively few and are all clustered at low inclination
angles. Nevertheless, the correlation is quite acceptable, as
witnessed by the orthogonal distances and the coefficients of the
regression line. 
It could be expected that two methods that rely on similar principles
would give results closer to each other than methods based on
different principles. 
We thus found it particularly gratifying to note the good agreement between
either of our two methods and the kinematics, which is considered to give
very reliable determinations and relies on very different assumptions.

\begin{center}
\begin{table}
\caption{ Weighted correlation coefficients for the PA correlations 
(upper triangle) and $cos($IA$)$ (lower triangle).} 
\label{cortab}
\begin{tabular}{rrrrrrrr}\\
\hline
      & 1 st   &   2nd  &   RC3 & G     & D      &    K   & P  \\\hline
 1st  &        & $0.92$ & $0.94$& $0.87$& $0.94$ & $0.96$& $0.79$ \\
 2nd  & $0.98$ &        & $0.96$& $0.89$& $0.96$ & $0.92$& $0.83$ \\
 RC3  & $0.87$ & $0.88$ &       & $0.96$& $0.98$ & $0.88$& $0.89$ \\
 G    & $0.89$ & $0.77$ & $0.84$&       & $0.89$ & $0.79$& $0.75$ \\
 D    & $0.85$ & $0.91$ & $0.94$& $0.51$&        & $0.94$& $0.59$ \\
 K    & $0.98$ & $0.85$ & $0.93$& $0.86$& $0.89$ &       & $0.81$ \\
 P    & $0.77$ & $0.69$ & $0.83$& $0.68$& $0.82$ & $0.71$&     \\\hline
\end{tabular}
\end{table}
\end{center}

One can thus conclude that all methods are acceptable for statistical
analysis of samples of disc galaxies. In particular our two methods
are as good as the other methods used so far and can give equally
reliable estimates of the deprojection parameters. On the other hand
if we are interested in the PA and IA of a particular galaxy it is
best to apply several methods. The reason is that the different
methods suffer from different
biases - i.e. warps in the external parts, non elliptical isophotes
or not well defined backgrounds - which might be more or less
important in a particular case. 

\begin{center}
\begin{table}
\caption{ Zero points of the best fitting straight line for the PA correlations 
(upper triangle) and $cos($IA$)$ (lower triangle).} 
\label{atab}
\begin{tabular}{rrr@{\hspace{-.005mm}}r
@{\hspace{-.005mm}}r@{\hspace{-.005mm}}r@{\hspace{-.005mm}}r
@{\hspace{-.005mm}}r}\\\hline
      & 1 st   &   2nd  &   RC3 & G     & D      &    K   & P  \\\hline
 1st  &        & $7.2$ & $12.2$& $5.7$& $8.5$ & $9.9$& $5.2$ \\
 2nd  & $7.2$ &        & $3.0$& $-0.1$& $1.0$ & $-4.1$& $-4.8$ \\
 RC3  & $-8.8$ & $-19.0$ &       & $4.1$& $-1.4$ & $-8.2$& $4.5$ \\
 G    & $-3.1$ & $9.3$ & $-19.5$&       & $14.6$ & $1.2$& $15.6$ \\
 D    & $9.5$ & $-8.5$ & $17.1$& $7.3$&        & $-2.0$& $-2.5$ \\
 K    & $5.3$ & $-6.9$ & $22.4$& $-10.2$& $15.5$ &       & $-26.5$ \\
 P    & $8.7$ & $4.3$ & $13.3$& $29.2$& $-12.7$ & $7.7$&     \\\hline
\end{tabular}
\end{table}
\end{center}

\begin{center}
\begin{table}
\caption{Slopes of the best fitting straight line for the PA correlations 
(upper triangle) and $cos($IA$)$  (lower triangle).} 
\label{btab}
\begin{tabular}{rrrrrrrr}\\
\hline
      & 1 st   &   2nd  &   RC3 & G     & D      &    K   & P  \\\hline
 1st  &        & $0.96$ & $0.89$& $0.87$& $0.90$ & $0.87$& $0.97$ \\
 2nd  & $0.93$ &        & $0.95$& $1.01$& $0.97$ & $0.98$& $1.04$ \\
 RC3  & $1.02$ & $1.12$ &       & $0.94$& $1.01$ & $1.03$& $1.01$ \\
 G    & $0.99$ & $0.93$ & $1.12$&       & $0.85$ & $1.02$& $0.91$ \\
 D    & $0.88$ & $1.03$ & $0.79$& $0.88$&        & $1.02$& $1.03$ \\
 K    & $0.93$ & $1.04$ & $0.81$& $0.91$& $0.92$ &       & $1.30$ \\
 P    & $0.88$ & $0.91$ & $0.89$& $0.83$& $1.28$ & $0.91$&     \\\hline
\end{tabular}
\end{table}
\end{center}

\begin{acknowledgements}
In preparing this paper we made extensive use of the CDS Strasbourg database. 
CGG and CB acknowledge financial support by the Direcci\'on de
Investigaci\'on cint\'{\i}fica y T\'ecnica under contract PB97-0411
\end{acknowledgements}

\vfill
\pagebreak


\begin{thebibliography}{}
\bibitem[1989]{afanasiev}
Afanasiev, V.L., Sil'chenko, O.K., Zasov, A.V. 1989, AA, 213, L9
\bibitem[2000]{aguerri}
Aguerri, J.A.L., Mu\~noz-Tu\~n\'on, C., Varela, A.M., Prieto, M. 2000, AA
361, 841
\bibitem[1998]{alonso}
Alonso-Herrero, A., Simpson, C., Ward, M., Wilson, A.S. 1998, ApJ,
495, 196
\bibitem[1988]{arsenault}
Arsenault, R., Boulesteix, J., Georgelin, Y., Roy, J.-R. 1988, AA,
  200, 29
\bibitem[1993]{athanassoula}
Athanassoula, E., Garc\'{\i}a-G\'omez, C., Bosma, A. 1993, AAS, 102, 229
\bibitem[2002]{athanassoula2}
Athanassoula, E., Misiriotis, A. 2002, MNRAS in press and astro-ph/0111449
\bibitem[1985]{atherton}
Atherton, P.D., Reay, N.K., Taylor, K., 1985, MNRAS, 216, 17p
\bibitem[1973]{balkowski}
Balkowski, C. AA, 29, 43
\bibitem[1986]{baumgart}
Baumgart, Ch.W., Peterson, Ch.J. 1986, PASP, 98, 56
\bibitem[1987]{baldwin}
Baldwin, J.A., Wilson, A.S., Whittle, M., 1987, ApJ, 319, 84
\bibitem[1986]{ball}
Ball. R. 1986, ApJ, 307, 453
\bibitem[1988]{becker}
Becker, R., Mebold, U., Reif, K.,van Woerden, H. 1988, AA, 203, 21
\bibitem[1986]{beckman}
Beckman, J.E., Bransgrove, S.G., Phillips, J.P. 1986, AA, 157, 49
\bibitem[1987]{begeman1}
Begeman, K.G. 1987, Ph. D. Thesis.
\bibitem[1989]{begeman2}
Begeman, K.G. 1989, AA, 223, 47
\bibitem[1991]{begeman3}
Begeman, K.G., Broeils, A.H., Sanders, R.H. 1991, MNRAS, 249, 523
\bibitem[1989]{bergeron}
Bergeron, J., Petitjean, P., Durret, F., 1989, AA, 213, 61
\bibitem[1979a]{blackman1}
Blackman, C.P. 1979a, MNRAS, 186, 701
\bibitem[1979b]{blackman2}
Blackman, C.P. 1979b, MNRAS, 186, 717
\bibitem[1981]{blackman3}
Blackman, C.P. 1981, MNRAS, 195, 451
\bibitem[1982]{blackman4}
Blackman, C.P. 1982, MNRAS, 200, 407
\bibitem[1983]{blackman5}
Blackman, C.P. 1983, MNRAS, 202, 379
\bibitem[1984]{blackman6}
Blackman, C.P. 1984, MNRAS, 207, 9
\bibitem[1997]{bland}
Bland-Hawthorn, J., Gallimore, J.F., Tacconi, L.,J., Brinks, E., Baum, S.A., 
Antonucci, R.R.J., Cecil, G.N., ApSS, 248, 9
\bibitem[1981]{boroson}
Boroson, T. 1981, ApJS, 46, 177
\bibitem[1981]{bosma1}
Bosma, A. 1981, AJ, 86, 1791
\bibitem[1981]{bosma2}
Bosma, A., Goss, W.M., Allen, R.J. 1981, AA, 93, 106
\bibitem[1977]{bosma3}
Bosma, A., van der Hulst, J.M., Sullivan III, W.T. 1977, AA, 57, 373
\bibitem[1988]{bottema}
Bottema, R., 1988, AA 197, 105
\bibitem[1962]{burbidge1}
Burbidge, E.M., Burbidge, G.R. 1962, ApJ, 135, 366
\bibitem[1964]{burbidge2}
Burbidge, E.M., Burbidge, G.R. 1964, ApJ, 140, 1445
\bibitem[1968a]{burbidge3}
Burbidge, E.M., Burbidge, G.R. 1968a, ApJ, 151, 99
\bibitem[1968b]{burbidge4}
Burbidge, E.M., Burbidge, G.R. 1968b, ApJ, 154, 857
\bibitem[1959]{burbidge5}
Burbidge, E.M., Burbidge, G.R., Prendergast, K.H. 1959, ApJ, 130, 26
\bibitem[1960]{burbidge6}
Burbidge, E.M., Burbidge, G.R., Prendergast, K.H. 1960, ApJ, 132, 654
\bibitem[1961]{burbidge7}
Burbidge, E.M., Burbidge, G.R., Prendergast, K.H. 1961, ApJ, 134, 874
\bibitem[1962]{burbidge8}
Burbidge, E.M., Burbidge, G.R., Prendergast, K.H. 1963, ApJ, 138, 375
\bibitem[1987a]{buta1}
Buta, R. 1987a, ApJS, 64, 1
\bibitem[1987b]{buta2}
Buta, R. 1987b, ApJS, 64, 383
\bibitem[1988]{buta3}
Buta, R. 1988, ApJS, 66, 233
\bibitem[1995]{buta4}
Buta, R. 1995, ApJS, 96, 39
\bibitem[1998]{buta5}
Buta, R., Purcell, G.B. 1998, AJ, 115, 484
\bibitem[to 33]{buta6}
Buta, R., private comunication to 33
\bibitem[1969]{carranza}
Carranza, G., Crillon, R., Monnet, G. 1969, AA, 1, 479
\bibitem[1967a]{chincarini1}
Chincarini, G., Walker, M.F. 1967a, ApJ, 149, 487
\bibitem[1967b]{chincarini2}
Chincarini, G., Walker, M.F. 1967b, ApJ, 147, 407
\bibitem[1982]{comte1}
Comte, G., Duquennoy, A. 1982, AA, 114, 7
\bibitem[1979]{comte2}
Comte, G., Monnet, G., Rosado, M. 1979, AA, 72, 73
\bibitem[1982]{considere1}
Consid\`ere, S., Athanassoula, E. 1982, AA, 111, 28
\bibitem[1988]{considere2}
Consid\`ere, S., Athanassoula, E. 1988, AAS, 76, 365
\bibitem[1997]{corbelli}
Corbelli, E., Schneider, S.E., 1997, ApJ, 479, 244
\bibitem[1991]{corradi}
Corradi, R.L.M., Boulesteix, J., Bosma, A., Capaccioli, M., Amram, P.,
  Mar\-ce\-lin, M. 1991, AA 244, 27
\bibitem[1993]{courtes}
Court\'es, G., Petit, H., Hua, C.T., Martin, P., Blecha, A., Huguenin,
D., Golay, M. 1993, AA, 268, 419
\bibitem[1996]{crocker}
Crocker, D.A., Baugus, P.A., Buta, R. 1996, ApJSS, 105, 353
\bibitem[1942]{danver}
Danver, C.G. 1942, Lund. Obs. Ann. No. 10
\bibitem[1981]{dickson}
Dickson, R.J., Hodge, P.W. 1981, AJ, 86, 826
\bibitem[1994]{dejong}
de Jong, R.S., van der Kruit, P.C. 1994, AAS, 106, 451
\bibitem[1973]{devaucouleurs2}
de Vaucouleurs, G. 1973, ApJ, 181, 31
\bibitem[1959]{devaucouleurs3}
de Vaucouleurs, G., 1959, ApJ 170, 728
\bibitem[1991]{devaucouleurs}
de Vaucouleurs, G., de Vaucouleurs, A., Corwin, H.G., Bu\-ta, R.J.,
  Paturel, G., Fouqu\'e, P. 1991, Third Referen\-ce Catalogue of Bright
  Galaxies, Springer, New York. (RC3)
\bibitem[1987]{deul}
Deul, E.R., van der Hulst, J.M., 1987, AAS, 67, 509
\bibitem[1985]{duval}
Duval, M.F., Monnet, G. 1985, AAS, 61, 141
\bibitem[1987]{elmegreen}
Elmegreen, D.M., Elmegreen, B.G. 1987, ApJ, 314, 3
\bibitem[1996]{evans}
Evans, I.N., Koratkar, A.P., Storchi-Bergmann, T., Kirk\-patrick, H.,
  Heckman, T.M., Wilson, A.S. 1996, ApJSS, 105, 93
\bibitem[1997]{feinstein}
Feinstein, C. 1997, ApJSS, 112, 29
\bibitem[2001]{fridman}
Fridman, A.M., Khoruzhii, O.V., Lyakhovich, V.V., Sil'chen\-ko, O.K., Zasov, A.V.,
Afanasiev, V.L., Dodonov, S.N., Boulesteix, J. 2001, AA, 371, 538
\bibitem[1991]{gga1}
Garc\'{\i}a-G\'omez, C., Athanassoula, E. 1991, AAS, 89, 159 (GGA)
\bibitem[1993]{gga2}
Garc\'{\i}a-G\'omez, C., Athanassoula, E. 1993, AAS, 100, 431
\bibitem[1979]{goad}
Goad, J.W., De Veny, J.B., Goad, L.E. 1979, ApJS, 39, 439
\bibitem[1997]{gonzalez}
Gonz\'alez Delgado, R.M., P\'erez, E., Tadhunter, C., V\'{\i}lchez,
  J.M., Rodr\'{\i}guez Espinosa, J.M. 1997, ApJSS, 108, 155
\bibitem[1982]{gottesman}
Gottesman, S.T. 1982, AJ, 87, 751
\bibitem[1985]{grosbol1}
Grosb\o l, P.J. 1985, AAS, 60, 261
\bibitem[1998]{grosbol2}
Grosb\o l, P.J., Patsis, P.A. 1998, AA, 336, 840
\bibitem[1988]{guhatakurta}
Guhatakurta, P., van Gorkom, J.H., Kotany, C.G., Balkowski, C.
  1988, AJ, 96, 851
\bibitem[1983]{hackwell}
Hackwell, J.A., Schweizer, F. 1983, ApJ, 265, 643
\bibitem[1995]{helfer}
Helfer, T.T., Blitz, L. 1995, ApJ, 450, 90
\bibitem[1996]{heraudeau}
H\'eraudeau, Ph., Simien, F. 1996, AAS, 118, 111
\bibitem[1990]{hodge}
Hodge, P.W., Gurwell, M., Goldader, J.D., Kennicutt, R.C.Jr. 1990,
  ApJSS, 73, 661
\bibitem[1999]{hodge2}
Hodge, P.W., Balsley, J., Wyder, T.K., Skelton, B.P. 1999, PASP, 111, 685
\bibitem[1999]{hunt}
Hunt, L.K., Malkan, M.A., Rush, B., Bicay, M.D., Nelson, B.O., Stanga,
R.M., Webb, W. 1999, ApJSS, 125, 349
\bibitem[1975]{hutchmeier1}
Hutchmeier, W.K. 1975, AA, 45, 259
\bibitem[1979]{hutchmeier2}
Hutchmeier, W.K., Witzel, A. 1979, AA 74, 138
\bibitem[1982]{iye}
Iye, M., Okamura, S., Hamabe, M., Watanabe, M., 1982, ApJ, 256, 103
\bibitem[1989]{kaneko1}
Kaneko, N., Morita, K., Fukui, Y., Sugitani, K., Iwata, T., Nakai, N.,
  Kaifu, N., Listz, H. 1989, ApJ, 337, 691
\bibitem[1992]{kaneko3}
Kaneko, N., Satoh, T., Toyama, K., Sasaki, M., Nishima, M., Yamamoto, M., 
1992, AJ, 103, 422
\bibitem[1997]{kaneko2}
Kaneko, N., Aoki, K., Kosugi, G., Ohtani, H., Yoshida, M., Tomaya, K.,
  Satoh, T., Sasaki, M. 1997, AJ, 114, 94
\bibitem[1981]{kennicutt}
Kennicutt, R.C. 1981, AJ, 86, 1847
\bibitem[1997]{knapen1}
Knapen, J.H. 1997, MNRAS, 286, 403
\bibitem[1998]{knapen2}
Knapen, J.H. 1998, MNRAS, 297, 255
\bibitem[1993]{knapen3}
Knapen, J.H., Arnth-Jensen, N., Cepa, J., Beckman, J.E. 1993, AJ, 106, 56
\bibitem[1993]{knapen4}
Knapen, J.H., Cepa, J., Beckman, J.E., del Rio, S., Pedlar, A., 1993, ApJ, 416, 563
\bibitem[1993]{koribalski}
Koribalski, R., Dahlem, M., Mebold, U., Brinks, E. 1993, AA, 268,14
\bibitem[1998]{laine}
Laine, S., Gottesman, S.T. 1998, MNRAS, 297, 1041
\bibitem[1982]{lauberts}
Lauberts, A. 1982 in The ESO/Uppsala Survey of the ESO(B)
Atlas. Garching: European Southern Observatory
\bibitem[1995]{listz}
Listz, H., Dickey, J.M. 1995, AJ, 110, 998
\bibitem[1998]{lu}
Lu, N.Y. 1998, ApJ 506, 673
\bibitem[1998]{ma}
Ma, J., Peng, Q.-H., Gu, Q-S. 1998, AAS, 130, 449
\bibitem[1994]{marcelin}
Marcelin, M., Petrosian, A.R., Amram, P., Boulesteix, J. 1994, AA,
  282, 363
\bibitem[1996]{mathewson}
Mathewson, D.S., Ford, V.L. 1996, ApJS 107, 97
\bibitem[1984]{maucherat}
Maucherat, A.J., Dubout-Crillon, R., Monnet, G., Figon, P., 1984, AA,
133, 341
\bibitem[1984]{meyssonier}
Meyssonier, N. 1984, AAS, 58, 351
\bibitem[1981]{milliard}
Milliard, B., Marcelin, M. 1981, AA, 95, 59
\bibitem[1981]{monnet}
Monnet, G., Paturel, G., Simien, F. 1981, AA, 102, 119
\bibitem[1993]{mulder}
Mulder, P.S., van Driel, W. 1993, AA, 272, 63
\bibitem[1980]{newton}
Newton, K., 1980, MNRAS, 190, 689
\bibitem[1978]{okamura}
Okamura, S. 1978, PASJ, 30, 91
\bibitem[1989]{ondrechen}
Ondrechen, M.P., van der Hulst, J.M., Hummel, E. 1989, ApJ, 342, 39
\bibitem[1993]{oosterloo}
Oosterloo, T., Shostak, S. 1993, AAS, 99, 379
\bibitem[1999]{peletier}
Peletier, R.F., Knapen, J.H., Shlosman, I., P\'erez-Ram\'{\i}rez, Nadeau, D.,
Doyon, R., Rodr\'{\i}guez Espinosa, P\'erez Garc\'{\i}a, 1999, ApJS, 125, 363
\bibitem[1982]{pellet}
Pellet, A., Simien, F. 1982, AA, 106, 214
\bibitem[1984]{pence1}
Pence, W.D., Blackman, C.P. 1984, MNRAS, 207, 9
\bibitem[1990]{pence2}
Pence, W.D., Taylor, K., Atherton, P. 1990, ApJ, 357, 415
\bibitem[1998]{petit1}
Petit, H. 1998, AAS, 131, 317
\bibitem[1996]{petit2}
Petit, H., Hua, C.T., Bersier, D., Court\'es, G. 1996, AA, 309, 446
\bibitem[1991]{planesas}
Planesas, P., Scoville, N., Myers, S.T. 1991, ApJ, 369, 364
\bibitem[1992]{numerical}
Press, W.H., Teukiolsky, S.A., Vetterling, W.H., Flannery, B.P. 1992, 
Numerical Recipes, Cambridge Univ. Press.
\bibitem[1992]{prieto}
Prieto, M., Longley, D.P.T., Perez, E., Beckman, E., Varela, A.M.,
  Cepa, J. 1992, AAS, 93, 557
\bibitem[1992]{puerari}
Puerari, I., Dottori, H.A. 1992, AAS, 93, 469
\bibitem[1970]{roberts}
Roberts, M.S., Warren, J.L. 1970, AA, 6, 165
\bibitem[1998]{rodrigues}
Rodrigues, I., Dottori, H., Cepa, J., Vilchez, J. 1998, AAS, 128, 545
\bibitem[1971]{rogstad}
Rogstad, D.H., Shostak, G.S., 1971, AA, 13, 99
\bibitem[1990]{rots}
Rots, A.H. Bosma, A., van der Hulst, J.M., Athanassoula, E., Crane,
  P.C. 1990, AJ, 100, 387
\bibitem[1996]{rozas1}
Rozas, M., Beckman, J.E., Knapen, J.H. 1996, AA, 307, 735
\bibitem[2000a]{rozas2}
Rozas, M., Zurita, A., Beckman, J.E. 2000a, AA, 354, 823
\bibitem[2000b]{rozas3}
Rozas, M., Zurita, A., Beckman, J.E., P\'erez, D., 2000b AAS, 142, 259
\bibitem[1999]{rozas4}
Rozas, M., Zurita, A., Heller, C.H., Beckman, J.E. 1999, AAS, 135, 145
\bibitem[1964]{rubin1}
Rubin, V.C., Burbidge, E.M., Burbidge, G.R., Prendergast, K.H. 1964,
  ApJ, 140, 80
\bibitem[1965]{rubin2}
Rubin, V.C., Burbidge, E.M., Burbidge, G.R., Crampin, D.J.,
  Prendergast, K.H. 1965, ApJ, 141, 759
\bibitem[1980]{rubin3}
Rubin, V.C., Ford, W.K.Jr, Thonnard, N. 1980, ApJ, 238, 471
\bibitem[1996]{ryder1}
Ryder, S.D., Buta, R.J., Toledo, H., Shukla, H., Staveley-smith, L.,
  Walsh, W. 1996, ApJ, 460, 665
\bibitem[1998]{ryder2}
Ryder, S.D., Zasov, A.V., McIntyre, Walsh, W., Sil'chenko, O.K. 1998,
  MNRAS, 293, 411
\bibitem[2000]{sanchez}
S\'anchez-Portal, M., D\'{\i}az, A.I., Terlevich, R., Terlevich, E.,
\'Alvarez-\'Alvarez, M., Aretxaga, I. 2000, MNRAS, 312, 2
\bibitem[2000]{schmitt}
Schmitt, M., Kinney, A.L., 2000, ApJSS, 128, 479
\bibitem[1992]{scowen}
Scowen, P.A., Dufour, R.J., Hester, J.J. 1992, AJ, 104, 92
\bibitem[1975]{shane}
Shane, W.W., 1975, in La dynamique des galaxies spirales,
  C.N.R.S. Int. Colloq. N. 241, Paris
\bibitem[1955]{stocke}
Stocke, J. 1955, AJ, 60, 216
\bibitem[1997]{thean}
Thean, A.H.C., Mundell, C.G., Pedlar, A., Nicholson, R.A. 1997,
  MNRAS, 290, 15
\bibitem[1995]{tsvetanov}
Tsvetanov, Z., Perosian, A. 1995, ApJSS, 101, 287
\bibitem[1974]{tully}
Tully, R.B. 1974, ApJS, 27, 437
\bibitem[1980]{vanalbada1}
van Albada, G.D. 1980, AA, 90, 123
\bibitem[1975]{vanalbada2}
van Albada, G.D., Shane, W.W. 1975, AA, 42, 433
\bibitem[1988]{vanderhulst}
van der Hulst, T., Sancisi, R. 1988, AJ, 95, 1354
\bibitem[1973]{vanderkruit1}
van der Kruit, P.C. 1973, ApJ, 186, 807
\bibitem[1974]{vanderkruit2}
van der Kruit, P.C. 1974, ApJ, 192, 1
\bibitem[1976]{vanderkruit3}
van der Kruit, P.C. 1976, AAS, 25, 527
\bibitem[1996]{vonlinden}
von Linden, S., Reuter, H.P., Heidt, J., Wielebinski, R., Pophl, M.
1996, AA, 315, 52
\bibitem[1973]{warner}
Warner, P.J., Wright, M.C.H., Baldwin, J.E., 1973, MNRAS, 163, 163
\bibitem[1973]{weliachew}
Weliachew, L. Gottesman, S.T. 1973, AA, 24, 59
\bibitem[2001]{weiner}
Weiner, B.J., Williams, T.B., van Gorkom, J.H., Sellwod, J.A. 2001,
ApJ, 546, 916
\bibitem[1986]{wevers}
Wevers, B.M.H.R., van der Kruit, P.C., Allen, R.J. 1986, AAS, 66,
  505
\bibitem[2000]{wilke}
Wilke, K., M\"ollenhoff, C., Matthias, M. 2000, AA, 361, 507
\bibitem[2000]{wong}
Wong, T., Blitz, L. 2000, ApJ, 540, 771
\bibitem[1987]{zasov}
Zasov, A.V., Sil'chencko, O.K. 1987, SvA Lett., 13, 186

\end{thebibliography}
\end{document}